\documentclass[conference]{IEEEtran}
\IEEEoverridecommandlockouts

\usepackage{graphicx}
\usepackage{caption}
\usepackage{amsmath}
\usepackage{balance}
\usepackage{cite}
\usepackage{url}
\usepackage{bbm}
\usepackage[table,dvipsnames]{xcolor}
\usepackage[font=small,subrefformat=parens,labelformat=parens]{subfig}

\def\BibTeX{{\rm B\kern-.05em{\sc i\kern-.025em b}\kern-.08em
    T\kern-.1667em\lower.7ex\hbox{E}\kern-.125emX}}
    
\newcommand{\blue}[1]{\textcolor{blue}{#1}}
\newcommand{\red}[1]{\textcolor{red}{#1}}

\usepackage{amsmath,amssymb}
\usepackage{xifthen}
\usepackage{mathtools}
\usepackage{enumerate}
\usepackage{microtype}
\usepackage{xspace}
\usepackage{bm}
\usepackage[T1]{fontenc}
\usepackage{fancyhdr}
\usepackage{lastpage}
\usepackage{bbm}

\newcommand{\mbs}[1]{\bm{#1}}
\newcommand{\vect}[1]{{\lowercase{\mbs{#1}}}}

\newcommand{\rv}[1]{{\mathsf{#1}}}
\newcommand{\rvVec}[1]{\pmb{\mathsf{#1}}}

\renewcommand{\Re}[1][]{\ifthenelse{\isempty{#1}}{\operatorname{Re}}{\operatorname{Re}\left(#1\right)}}
\renewcommand{\Im}[1][]{\ifthenelse{\isempty{#1}}{\operatorname{Im}}{\operatorname{Im}\left(#1\right)}}

\newcommand{\snr}{\mathsf{snr}}

\newcommand{\cv}{\vect{c}}

\newcommand{\ev}{\vect{e}}

\newcommand{\yv}{\vect{y}}

\newcommand{\Ac}{{\mathcal A}}

\newcommand{\Sc}{{\mathcal S}}

\newcommand{\EE}{\mathbb{E}}

\newcommand{\CN}[1][]{\ifthenelse{\isempty{#1}}{\mathcal{N}_{\mathbb{C}}}{\mathcal{N}_{\mathbb{C}}\left(#1\right)}}

\renewcommand{\P}[1][]{\ifthenelse{\isempty{#1}}{\mathbb{P}}{\mathbb{P}\left[{#1}\right]}}
\newcommand{\E}[1][]{\ifthenelse{\isempty{#1}}{\mathbb{E}}{\mathbb{E}\left[#1\right]}}
\newcommand{\I}[1][]{\ifthenelse{\isempty{#1}}{\mathbb{I}}{\mathbb{I}\left\{#1\right\}}}
\renewcommand{\det}[1][]{\ifthenelse{\isempty{#1}}{\mathrm{det}}{\mathrm{det}\left(#1\right)}}
\newcommand{\trace}[1][]{\ifthenelse{\isempty{#1}}{{\rm tr}}{\mathrm{tr}\left(#1\right)}}
\newcommand{\rank}[1][]{\ifthenelse{\isempty{#1}}{\mathrm{rank}}{\mathrm{rank}\left(#1\right)}}
\newcommand{\diag}[1][]{\ifthenelse{\isempty{#1}}{\mathrm{diag}}{\mathrm{diag}\left(#1\right)}}
\newcommand{\Cov}[1][]{\ifthenelse{\isempty{#1}}{\mathsf{Cov}}{\mathsf{Cov}\left(#1\right)}}

\newtheorem{remark}{Remark}
\newtheorem{theorem}{Theorem}
\newtheorem{lemma}{Lemma}

\newcounter{enumi_saved}
\usepackage{answers}
\Newassociation{solution}{Solution}{solutionfile}

\AtBeginDocument{\Opensolutionfile{solutionfile}[\jobname]}
\AtEndDocument{\Closesolutionfile{solutionfile}\clearpage
}

\IfFileExists{MinionPro.sty}{
}{
}

\usepackage[toc,acronym]{glossaries}
\glsdisablehyper
\newglossaryentry{wlog}
{
	name={w.l.o.g.},
	description={without loss of generality},
	first={\glsentrydesc{wlog} (\glsentrytext{wlog})}
}

\newglossaryentry{sic}
{
	name={SIC},
	description={successive interference cancellation},
	first={\glsentrydesc{sic} (\glsentrytext{sic})}
}

\newglossaryentry{plr}
{
	name={PLR},
	description={packet loss rate},
	first={\glsentrydesc{plr} (\glsentrytext{plr})}
}

\newglossaryentry{pupe}
{
	name={PUPE},
	description={per-user probability of error},
	first={\glsentrydesc{pupe} (\glsentrytext{pupe})}
}

\newglossaryentry{pmf}
{
	name={PMF},
	description={probability mass function},
	first={\glsentrydesc{pmf} (\glsentrytext{pmf})}
}

\newglossaryentry{bpsk}
{
	name={BPSK},
	description={binary phase shift keying},
	first={\glsentrydesc{bpsk} (\glsentrytext{bpsk})}
}

\newglossaryentry{bac}
{
	name={BAC},
	description={binary adder channel},
	first={\glsentrydesc{bac} (\glsentrytext{bac})}
}

\newglossaryentry{mac}
{
  name={MAC},
  description={multiple-access channel},
  first={\glsentrydesc{mac} (\glsentrytext{mac})}
}

\newglossaryentry{simo}
{
	name={SIMO},
	description={single-input multiple-output},
	first={\glsentrydesc{simo} (\glsentrytext{simo})}
}

\newglossaryentry{mmse}
{
  name={MMSE},
  description={minimum mean-square error},
  first={\glsentrydesc{mmse} (\glsentrytext{mmse})}
}

\newglossaryentry{na}
{
  name={NA},
  description={normal approximation},
  first={\glsentrydesc{na} (\glsentrytext{na})},
  plural={NAs},
  descriptionplural={normal approximations},
  firstplural={\glsentrydescplural{na} (\glsentryplural{na})}
}

\newglossaryentry{zf}
{
  name={ZF},
  description={zero-forcing},
  first={\glsentrydesc{zf} (\glsentrytext{zf})}
}

\newglossaryentry{rzf}
{
  name={RZF},
  description={regularized zero-forcing},
  first={\glsentrydesc{rzf} (\glsentrytext{rzf})}
}

\newglossaryentry{sir}
{
  name={SIR},
  description={signal-to-interference ratio},
  first={\glsentrydesc{sir} (\glsentrytext{sir})}
}

\newglossaryentry{mse}
{
  name={MSE},
  description={mean-square error},
  first={\glsentrydesc{mse} (\glsentrytext{mse})}
}
\newglossaryentry{mmmse}
{
  name={M-MMSE},
  description={multicell MMSE},
  first={\glsentrydesc{mmmse} (\glsentrytext{mmmse})}
}

\newglossaryentry{ls}
{
  name={LS},
  description={least-squares},
  first={\glsentrydesc{ls} (\glsentrytext{ls})}
}

\newglossaryentry{mr}
{
  name={MR},
  description={maximum ratio},
  first={\glsentrydesc{mr} (\glsentrytext{mr})}
}

\newglossaryentry{ue}
{
  name={UE},
  description={user equipment},
  first={\glsentrydesc{ue} (\glsentrytext{ue})},
  plural={UEs},
  descriptionplural={user equipments},
  firstplural={\glsentrydescplural{ue} (\glsentryplural{ue})}
}

\newglossaryentry{ap}
{
  name={AP},
  description={access point},
  first={\glsentrydesc{ap} (\glsentrytext{ap})},
  plural={APs},
  descriptionplural={access points},
  firstplural={\glsentrydescplural{ap} (\glsentryplural{ap})}
}

\newglossaryentry{cpu}
{
  name={CPU},
  description={central processing unit},
  first={\glsentrydesc{cpu} (\glsentrytext{cpu})},
  plural={CPUs},
  descriptionplural={central processing units},
  firstplural={\glsentrydescplural{cpu} (\glsentryplural{cpu})}
}

\newglossaryentry{bs}
{
  name={BS},
  description={base station},
  first={\glsentrydesc{bs} (\glsentrytext{bs})},
  plural={BSs},
  descriptionplural={base stations},
  firstplural={\glsentrydescplural{bs} (\glsentryplural{bs})}
}

\newglossaryentry{ostbc}
{
  name={OSTBC},
  description={orthogonal space-time block code},
  first={\glsentrydesc{ostbc} (\glsentrytext{ostbc})},
  plural={OSTBCs},
  descriptionplural={orthogonal space-time block codes},
  firstplural={\glsentrydescplural{ostbc} (\glsentryplural{ostbc})}
}

\newglossaryentry{biawgn}
{
  name={bi-AWGN},
  description={binary-input additive white Gaussian noise},
  first={\glsentrydesc{biawgn} (\glsentrytext{biawgn})}
}

\newglossaryentry{flnf}
{
  name={FLNF},
  description={fixed-length no-feedback},
  first={\glsentrydesc{flnf} (\glsentrytext{flnf})}
}

\newglossaryentry{crc}
{
  name={CRC},
  description={cyclic redundancy check},
  first={\glsentrydesc{crc} (\glsentrytext{crc})}
}

\newglossaryentry{bsc}
{
  name={BSC},
  description={binary symmetric channel},
  first={\glsentrydesc{bsc} (\glsentrytext{bsc})}
  plural={BSCs},
  descriptionplural={binary symmetric channels},
  firstplural={\glsentrydescplural{bsc} (\glsentryplural{bsc})}
}

\newglossaryentry{bec}
{
  name={BEC},
  description={binary-erasure channel},
  first={\glsentrydesc{bec} (\glsentrytext{bec})}
}
\newglossaryentry{awgn}
{
  name={AWGN},
  description={additive white Gaussian noise},
  first={\glsentrydesc{awgn} (\glsentrytext{awgn})}
}
\newglossaryentry{3gpp}
{
  name={3GPP},
  description={3rd generation partnership project},
  first={\glsentrydesc{3gpp} (\glsentrytext{3gpp})}
}

\newglossaryentry{dl}
{
  name={DL},
  description={downlink},
  first={\glsentrydesc{dl} (\glsentrytext{dl})}
}

\newglossaryentry{LED}
{
  name={LED},
  description={light emitting diode},
  first={\glsentrydesc{LED} (\glsentrytext{LED})},
  plural={LEDs},
  descriptionplural={light emitting diodes},
  firstplural={\glsentrydescplural{LED} (\glsentryplural{LED})}
}

\newglossaryentry{lte}
{
  name={LTE},
  description={long term evolution},
  first={\glsentrydesc{lte} (\glsentrytext{lte})}
}
\newglossaryentry{ofdm}
{
  name={OFDM},
  description={orthogonal frequency-division multiplexing},
  first={\glsentrydesc{ofdm} (\glsentrytext{ofdm})}
}

\newglossaryentry{ul}
{
  name={UL},
  description={uplink},
  first={\glsentrydesc{ul} (\glsentrytext{ul})}
}
\newglossaryentry{rb}
{
  name={RB},
  description={resource block},
  first={\glsentrydesc{rb} (\glsentrytext{rb})},
  plural={RBs},
  descriptionplural={resource blocks},
  firstplural={\glsentrydescplural{rb} (\glsentryplural{rb})}
}
\newglossaryentry{cu}
{
  name={CU},
  description={channel use},
  first={\glsentrydesc{cu} (\glsentrytext{cu})},
  plural={CUs},
  descriptionplural={channel uses},
  firstplural={\glsentrydescplural{cu} (\glsentryplural{cu})}
}
\newglossaryentry{tti}
{
  name={TTI},
  description={transmission time interval},
  first={\glsentrydesc{tti} (\glsentrytext{tti})},
  plural={TTI's},
  descriptionplural={transmission time intervals},
  firstplural={\glsentrydescplural{tti} (\glsentryplural{tti})}
}
\newglossaryentry{iid}
{
  name={i.i.d.},
  description={independent and identically distributed},
  first={\glsentrydesc{iid} (\glsentrytext{iid})}
}
\newglossaryentry{psd}
{
  name={PSD},
  description={power spectral density},
  first={\glsentrydesc{psd} (\glsentrytext{psd})}
}

\newglossaryentry{mimo}
{
  name={MIMO},
  description={multiple-input multiple-output},
  first={\glsentrydesc{mimo} (\glsentrytext{mimo})}
}

\newglossaryentry{bler}
{
  name={BLER},
  description={block error probability},
  first={\glsentrydesc{bler} (\glsentrytext{bler})}
}
\newglossaryentry{mtc}
{
  name={MTC},
  description={machine-type communication},
  first={\glsentrydesc{mtc} (\glsentrytext{mtc})}
}
\newacronym{csi}{CSI}{channel state information}
\newglossaryentry{dvbt}
{
  name={DVB-T},
  description={terrestrial digital video broadcasting},
  first={\glsentrydesc{dvbt} (\glsentrytext{dvbt})}
}
\newglossaryentry{pat}
{
  name={PAT},
  description={pilot-assisted transmission},
  first={\glsentrydesc{pat} (\glsentrytext{pat})}
}
\newglossaryentry{ml}
{
  name={ML},
  description={maximum likelihood},
  first={\glsentrydesc{ml} (\glsentrytext{ml})}
}
\newglossaryentry{dt}
{
  name={DT},
  description={dependence-testing},
  first={\glsentrydesc{dt} (\glsentrytext{dt})}
}
\newglossaryentry{ustm}
{
  name={USTM},
  description={unitary space-time modulation},
  first={\glsentrydesc{ustm} (\glsentrytext{ustm})}
}
\newglossaryentry{siso}
{
  name={SISO},
  description={single-input single-output},
  first={\glsentrydesc{siso} (\glsentrytext{siso})}
}
\newglossaryentry{snn}
{
  name={SNN},
  description={scaled nearest-neighbor},
  first={\glsentrydesc{snn} (\glsentrytext{snn})}
}

\newglossaryentry{snnd}
{
  name={SNND},
  description={scaled nearest-neighbor decoding},
  first={\glsentrydesc{snnd} (\glsentrytext{snnd})}
}
\newglossaryentry{rcus}
{
  name={RCUs},
  description={random-coding union bound with parameter $s$},
  first={\glsentrydesc{rcus} (\glsentrytext{rcus})}
}
\newglossaryentry{osd}
{
  name={OSD},
  description={ordered statistics decoding},
  first={\glsentrydesc{osd} (\glsentrytext{osd})}
}
\newglossaryentry{llr}
{
  name={LLR},
  description={log-likelihood ratio},
  first={\glsentrydesc{llr} (\glsentrytext{llr})}
   plural={LLR's},
  descriptionplural={log-likelihood ratios},
  firstplural={\glsentrydescplural{llr} (\glsentryplural{llr})}
}
\newglossaryentry{qpsk}
{
  name={QPSK},
  description={quaternary phase shift keying},
  first={\glsentrydesc{qpsk} (\glsentrytext{qpsk})}
}
\newglossaryentry{nlos}
{
  name={NLOS},
  description={non-line-of-sight},
  first={\glsentrydesc{nlos} (\glsentrytext{nlos})}
}
\newglossaryentry{los}
{
  name={LOS},
  description={line-of-sight},
  first={\glsentrydesc{los} (\glsentrytext{los})}
}
\newglossaryentry{mgf}
{
  name={MGF},
  description={moment-generating function},
  first={\glsentrydesc{mgf} (\glsentrytext{mgf})}
}
\newglossaryentry{cgf}
{
  name={CGF},
  description={cumulant-generating function},
  first={\glsentrydesc{cgf} (\glsentrytext{cgf})},
  plural={CGFs},
  descriptionplural={cumulant-generating functions},
  firstplural={\glsentrydescplural{cgf} (\glsentryplural{cgf})}
}
\newglossaryentry{pdf}
{
  name={PDF},
  description={probability density function},
  first={\glsentrydesc{pdf} (\glsentrytext{pdf})},
   plural={PDF's},
  descriptionplural={probability density functions},
  firstplural={\glsentrydescplural{pdf} (\glsentryplural{pdf})}
}

\newglossaryentry{ccdf}
{
  name={CCDF},
  description={complementary cumulative distribution function},
  first={\glsentrydesc{ccdf} (\glsentrytext{ccdf})}
}

\newglossaryentry{cdf}
{
  name={CDF},
  description={cumulative distribution function},
  first={\glsentrydesc{cdf} (\glsentrytext{cdf})}
}

\newglossaryentry{gmi}
{
  name={GMI},
  description={generalized mutual information},
  first={\glsentrydesc{gmi} (\glsentrytext{gmi})}
}

\newglossaryentry{arq}
{
  name={ARQ},
  description={automatic repeat request},
  first={\glsentrydesc{arq} (\glsentrytext{arq})}
}

\newglossaryentry{harq}
{
  name={HARQ},
  description={hybrid automatic repeat request},
  first={\glsentrydesc{harq} (\glsentrytext{harq})}
}

\newglossaryentry{fbl}
{
  name={FBL-NF},
  description={fixed blocklength with no feedback},
  first={\glsentrydesc{fbl} (\glsentrytext{fbl})}
}
\newglossaryentry{ssnf}
{
  name={SS-NF},
  description={single-shot no-feedback},
  first={\glsentrydesc{ssnf} (\glsentrytext{ssnf})}
}

\newacronym{urllc}{URLLC}{ultra-reliable low-latency communications}

\newglossaryentry{vlsf}
{
  name={VLSF},
  description={Variable-length stop-feedback},
  first={\glsentrydesc{vlsf} (\glsentrytext{vlsf})}
}

\newglossaryentry{vlnsf}
{
  name={VLNSF},
  description={variable-length noisy stop feedback},
  first={\glsentrydesc{vlnsf} (\glsentrytext{vlnsf})}
}

\newglossaryentry{dmt}
{
  name={DMT},
  description={diversity-multiplexing tradeoff},
  first={\glsentrydesc{dmt} (\glsentrytext{dmt})}
}

\newglossaryentry{dmdt}
{
  name={DMDT},
  description={diversity-multiplexing-delay tradeoff},
  first={\glsentrydesc{dmdt} (\glsentrytext{dmdt})}
}

\newglossaryentry{tddt}
{
  name={TDDT},
  description={throughput-diversity-delay tradeoff},
  first={\glsentrydesc{tddt} (\glsentrytext{tddt})}
}

\newglossaryentry{tdd}
{
  name={TDD},
  description={time-division duplex},
  first={\glsentrydesc{tdd} (\glsentrytext{tdd})}
}

\newglossaryentry{stbc}
{
  name={STBC},
  description={space-time block-codes},
  first={\glsentrydesc{stbc} (\glsentrytext{stbc})},
  plural={STBC's},
  descriptionplural={space-time block-codes},
  firstplural={\glsentrydescplural{stbc} (\glsentryplural{stbc})}
}

\newglossaryentry{harqir}
{
  name={HARQ-IR},
  description={hybrid automatic repeat request with incremental redundancy},
  first={\glsentrydesc{harqir} (\glsentrytext{harqir})}
}

\newglossaryentry{harqcc}
{
  name={HARQ-CC},
  description={hybrid automatic repeat request with chase combining},
  first={\glsentrydesc{harqcc} (\glsentrytext{harqcc})}
}

\newglossaryentry{wp1}
{
  name={w.p.1.},
  description={with probability one},
  first={\glsentrydesc{wp1} (\glsentrytext{wp1})}
}

\newglossaryentry{vlff}
{
  name={VLFF},
  description={variable-length full feedback},
  first={\glsentrydesc{vlff} (\glsentrytext{vlff})}
}

\newglossaryentry{dmc}
{
  name={DMC},
  description={discrete memoryless channel},
  first={\glsentrydesc{dmc} (\glsentrytext{dmc})}
  plural={DMCs},
  descriptionplural={discrete memoryless channels},
  firstplural={\glsentrydescplural{dmc} (\glsentryplural{dmc})}
}

\newglossaryentry{ssa}
{
	name={SSA},
	description={segmented slotted ALOHA},
	first={\glsentrydesc{ssa} (\glsentrytext{ssa})}
}

\newacronym{IRSA}{IRSA}{irregular repetition slotted ALOHA}
\newacronym{MPR}{MPR}{multi-packet reception}
\newacronym{SPR}{SPR}{single-packet reception}
\newacronym{IoT}{IoT}{internet of things}
\newacronym{LDPC}{LDPC}{low-density parity-check}

\usepackage{pgfplots}
\pgfplotsset{compat=1.14}
\usepgfplotslibrary{groupplots}
\usetikzlibrary{pgfplots.groupplots} 

\pgfplotsset{minor grid style={dotted}}
\pgfplotsset{major grid style={dashed}}
\usepackage{tikz}
\usetikzlibrary{fit,positioning,arrows,shapes,shapes.multipart,calc,arrows.meta,shapes.geometric,shapes.misc}
\usetikzlibrary{decorations.pathreplacing,angles,quotes,calligraphy}
\usetikzlibrary{fit,positioning,arrows,shapes,shapes.multipart,calc,arrows.meta}

\usepackage{pgfplots}
\pgfplotsset{compat=newest}
\usepgfplotslibrary{fillbetween}

\newcommand{\of}[1]{^{(#1)}}

\newcommand{\Bino}{{\rm Bino}}

\newcommand{\sub}[1]{_{\rm #1}}

\newcommand{\prob}[1]{\ensuremath{\mathbb{P}\left[#1\right]}}

\newcommand{\lrb}[1]{\left \lbrace {#1} \right \rbrace}															

\newcommand{\ind}[1]{{\mathbbm{1}{\left\{#1\right\}}}}

\makeatletter
\newcommand{\vast}{\bBigg@{4}}
\newcommand{\Vast}{\bBigg@{5}}
\makeatother

\newcommand{\revise}[1]{#1} 

\makeatletter
\newcommand{\removelatexerror}{\let\@latex@error\@gobble}
\makeatother

\title{Irregular Repetition Slotted ALOHA \\ Over the Binary Adder Channel} 

\author{
	\IEEEauthorblockN{Khac-Hoang Ngo, Alexandre Graell i Amat, and Giuseppe Durisi} 
	\IEEEauthorblockA{Department of Electrical Engineering, Chalmers University of Technology, 41296 Gothenburg, Sweden}
	\thanks{This project has received funding from the European Union's Horizon 2020 research and innovation programme under the Marie Skłodowska-Curie Grant Agreement No 101022113 and under Grant Agreement No 856967, and from the Swedish Research Council under grant 2021-04970.}
	\vspace{-.3cm}
}

\IEEEoverridecommandlockouts
\begin{document}
	
	\maketitle
	\begin{abstract} 
		We propose an irregular repetition slotted ALOHA~(IRSA) based random-access protocol for the binary adder channel~(BAC). The~BAC captures important physical-layer concepts, such as packet generation, per-slot decoding, and information rate, which are neglected in the commonly considered collision channel model. We divide a frame into slots and let users generate a packet, to be transmitted over a slot, from a given codebook. In a state-of-the-art scheme proposed by Paolini \emph{et al.} (2022), the codebook is constructed as the parity-check matrix of a BCH code. Here, we construct the codebook from independent and identically distributed binary symbols to obtain a random-coding achievability bound. 
	Our per-slot decoder progressively discards incompatible codewords from a list of candidate codewords, and can be improved by shrinking this list across iterations. In a regime of practical interests, 
		our scheme can resolve more colliding users in a slot and thus achieves a higher average sum rate than the scheme in Paolini \emph{et al.} (2022).
	\end{abstract}
		%
	
	\section{Introduction}\label{sec:intro}
	In recent years, modern random-access protocols~\cite{Berioli2016NOW} have been developed based on packet-based coding 
	at the users and \gls{sic} decoding at the receiver. In particular, \gls{IRSA}~\cite{Liva2011IRSA} employs packet repetitions: active users transmit multiple copies of their packets in randomly chosen slots of a fixed-length frame. This creates time diversity and boosts the throughput with respect to slotted ALOHA. 
	Modern random-access protocols have been adopted in commercial satellite communication systems, e.g., DVB-RCS2~\cite{ETSI2020DVB}. They provide a practical mean to achieve uncoordinated medium access in internet-of-things~(IoT) applications~\cite{Clazzer20195g}.
	Under an ``unsourced'' model where all users use the same codebook and the receiver aims to return an unordered list of messages, slotted ALOHA and \gls{IRSA} exhibit a large energy-efficiency gap to random-coding achievability bounds~\cite{PolyanskiyISIT2017massive_random_access,Ngo2022arXivUnsourced}. This gap can be reduced by combining \gls{IRSA} with more advanced multi-user code constructions and multi-user joint decoding; see, e.g.,~\cite{Vem2019,Marshakov2019Polar}.
	
	Most works on \gls{IRSA} assume a collision channel where successful decoding is achieved only in slots containing a single packet.  
	This model enables a direct connection between \gls{sic} decoding of \gls{IRSA} and graph-based iterative erasure decoding of \gls{LDPC} codes. This yields in turn a tractable analysis of the \gls{plr} and leads to the definition of a decoding threshold, i.e., the maximum average number of active users per slot such that vanishing \gls{plr} can be achieved as the number of slots per frame goes to infinity. However, this simple model does not capture some physical-layer concepts such as how a packet is generated, how packet decoding is performed in a slot, and the information rate 
    that can be achieved with vanishing \gls{plr}. 
	
	The \gls{bac}, where users transmit binary symbols and the receiver observes the noiseless sum of these symbols, provides a  more realistic setting to analyze \gls{IRSA} than the collision channel. The \gls{bac} resembles the interference-limited regime where noise is negligible, and captures the  aforementioned physical-layer concepts. Insights on designing codes for the two-user unsourced \gls{bac} were reported in~\cite{Liva2021codingUnsourced}. In~\cite{David1993adderchannel}, Bar-David, Plotnik, and Rom proposed a coding scheme, which we shall refer to as BPR, capable of correctly decoding $T$ colliding codewords 
	with $T$ known \emph{a priori}. The codebook therein is constructed as the parity-check matrix of a $T$-error-correcting binary Bose–Chaudhuri–Hocquenghem~(BCH) code. In~\cite{Paolini2022IRSA_BAC}, Paolini \emph{et al.} considered \gls{IRSA} over the \gls{bac} and let users generate packets from the BPR codebook. A unit pilot symbol, prepended to each codeword, allows the receiver to infer the number of active users in a slot. 
	The per-slot decoder can resolve collisions of up to $T$ users. The decoding threshold of this scheme follows from that of \gls{IRSA} with \gls{MPR}~\cite{Ghanbarinejad2013Irregular}. 
	
	In this paper, we propose a new \gls{IRSA}-based scheme for the \gls{bac}. As in~\cite{Paolini2022IRSA_BAC}, a unit pilot symbol is prepended to each codeword. However, different from~\cite{Paolini2022IRSA_BAC}, we generate the remaining symbols in an \gls{iid} manner from a Bernoulli distribution. This results in a random-coding achievability bound. Our per-slot decoder progressively discards incompatible codewords from a candidate list.
	This leads to better \gls{MPR} capability and higher asymptotic achievable average sum rate than the scheme in~\cite{Paolini2022IRSA_BAC} in the regime where the number of bits per user increases slowly with the number of users and the total number of bits normalized by the slot length is close to~$1$. Our decoder can be further improved by shrinking the candidate list across iterations. 
		
	\subsubsection*{Notation} Lowercase boldface letters denote vectors. We denote random quantities with nonitalic sans-serif letters, e.g., a scalar $\rv{x}$ and a vector $\rvVec{x}$, and  
	deterministic quantities 
	with italic letters, e.g., a scalar $x$ and a vector $\bm{x}$. 
	Calligraphic uppercase letters denote sets, 
	$[m:n]$ stands for the set $\{m,m+1,\dots,n\}$, 
	$[n] = [1:n]$, and 
	$\ind{\cdot}$ is the indicator function. 
	The Binomial distribution with parameters $(n,p)$ is denoted by $\mathrm{Bino}(n,p)$. 
	
	\section{System Model} \label{sec:model}
	
	\subsection{Random-Access Over the Binary Adder Channel} \label{sec:BAC}
	We consider a scenario with $K$ users, each active with probability $\mu$. Hence, the number of active users, $\rv{K}\sub{a}$, follows a $\mathrm{Bino}(K,\mu)$ distribution. The active users send encoded messages to a receiver over $n$ uses of a stationary memoryless \gls{bac}. 
	Let~$\rvVec{x}_k$ be the length-$n$ binary coded sequence transmitted by user~$k$. The received signal is the noiseless sum of the transmitted sequences, i.e., $\rvVec{y} \!=\! \sum_{k=1}^{\rv{K}\sub{a}} \rvVec{x}_k$.
	The receiver does not know $\rv{K}_{\rm a}$ {\em a priori} 
    and aims to recover all transmitted messages. 

	\subsection{Irregular Repetition Slotted ALOHA} \label{sec:IRSA}
	In~\gls{IRSA}, each length-$n$ frame is divided into $N$ slots and the users are frame- and slot-synchronous. A message is encoded in a packet (also called codeword) transmitted over a slot. An active user in a frame sends~$\rv{L}$ identical replicas of its packet in~$\rv{L}$ slots chosen uniformly without replacement. Here, $\rv{L}$ is called the degree of the transmitted packet or the transmitting user. It follows a probability distribution $\{\Lambda_L\}$ with $\Lambda_L = \prob{\rv{L} = L}$, which we write using a polynomial notation as $\Lambda(x) = \sum_{L} \Lambda_L x^L$. \gls{IRSA} was first proposed for the {collision channel}, in which slots containing a single packet (called singleton slots) always lead to successful decoding, whereas slots containing multiple packets lead to decoding failures~\cite{Liva2011IRSA}. 
    \revise{For a packet $\cv$, both the users and the receiver employ a common function $h(\cv)$ to generate the position of the replicas. See~\cite[App.]{Paolini2022IRSA_BAC} for details.}
	The receiver employs a \gls{sic} decoder that operates as follows. It seeks a singleton slot, decodes the packet therein, then locates 
    and removes	its replicas. These steps are repeated until no more singleton slots can be found.
	
	Consider a bipartite graph where variable nodes~(VNs) represent users, check nodes~(CNs) represent slots, and VN~$k$ is connected to CN~$s$ by an edge if user~$k$ transmits in slot~$s$. 
	Let $\lambda_L$ 
	be the probability that an edge is connected to a degree-$L$ VN. 
	We also use the polynomial notation 
	$\lambda(x) = \sum_L \lambda_L x^{L-1}$.  
	It holds that $\lambda(x) = \frac{\Lambda'(x)}{\Lambda'(1)}$~\cite{Richardson2008modern}, where $\Lambda'(x)$ is the first-order derivative of $\Lambda(x)$. 
	\gls{sic} decoding can be interpreted as erasure decoding on this bipartite graph: an edge connected to a VN is revealed if at least another edge connected to the same VN has been revealed; an edge connected to a CN is revealed if all other edges connected to the same CN have been revealed. 

	\section{An IRSA-Based Scheme for the BAC} \label{sec:IRSA_BAC}
	
	For the \gls{bac}, one can divide a length-$n$ frame into $N$ slots (assuming that $n$ divides $N$) and directly apply the original \gls{IRSA} protocol. 
    Define $G = \E[\rv{K}\sub{a}]/N = \mu K/N$ as the {channel load}.
	According to~\cite[Sec.~IV]{Liva2011IRSA}, as $K \to \infty$ and $N \to \infty$ while $G$ is fixed,  the \gls{plr} drops at a certain threshold value as the channel load $G$ increases. That is, all but a vanishing fraction of the transmitted messages are successfully decoded if the channel load is below a {decoding threshold}. The decoding threshold $G^*(\Lambda)$ is the maximum value of $G$ such that 
	$
	p > \lambda(1-e^{-pG\Lambda'(1)}),  \forall p \in (0,1]. 
	$
	We next propose a more advanced \gls{IRSA}-based scheme based on random coding and \gls{MPR}. 
	
	\subsection{Encoder}
	Let $n_0 + 1 = n/N$ be the slot length. We generate a codebook $\{\cv_m\}_{m=1}^M$ ($M \le 2^{n_0}$) where each codeword contains a unit pilot symbol followed by $n_0$ \gls{iid} binary symbols, each being~$1$ with probability~$\nu$. (The choice of $\nu$ will be discussed in Remark~\ref{rem:trade-off_nu}.)
	An active user selects a codeword and transmits~$\rv{L}$ copies, \revise{with $\rv{L} \sim \Lambda(x)$},  of this codeword in $\rv{L}$ slots chosen from the $N$ available slots. 
	We next propose two \gls{sic} decoders.
	
	\subsection{Erasure Decoding With Multi-Packet Reception (ED-MPR)} 
	\label{sec:PD_MPR}
	This decoder can resolve collisions in nonsingleton slots. In each iteration, it goes through all slots. 
	Consider slot $s$. \revise{The number of active users in this slot, $\rv{K}_s$, is called the slot degree. In the first channel use, the channel output is identical to $\rv{K}_s$, thus the receiver estimates perfectly $\rv{K}_s$.} 
 The receiver forms a list $\Sc_s$ of candidate codewords containing all codewords that could have been transmitted in the slot.\footnote{The candidate lists $\{\Sc_s\}_{s \in [N]}$ can be computed once and then stored at the receiver. \revise{Storing $\{\Sc_s\}_{s \in [N]}$, however, requires a large memory if $M$ is large.} Alternatively, the bipartite graph, and thus $\{\Sc_s\}_{s \in [N]}$, can be constructed by the receiver and communicated to the users as assumed in~\cite{Paolini2017GIRSA}. This is possible in systems where users go through an access procedure to join the network, after which the user population is known to the receiver. \revise{In this case, the receiver can impose a graph structure that allows for efficient storage.}} That is,
		$\Sc_s = \{i\in [M] \colon s \in h(\cv_i)\}$. 
	Next, the receiver attempts to resolve the slot by using a per-slot decoder with \gls{MPR} capability, which we describe in the next paragraph. If the slot is resolved, i.e., all transmitted codewords are decoded, the receiver removes the replicas of these codewords. These per-slot-decoding and interference-cancellation steps are repeated until no more resolvable slots can be found. 

	\subsubsection*{Per-Slot Decoder}
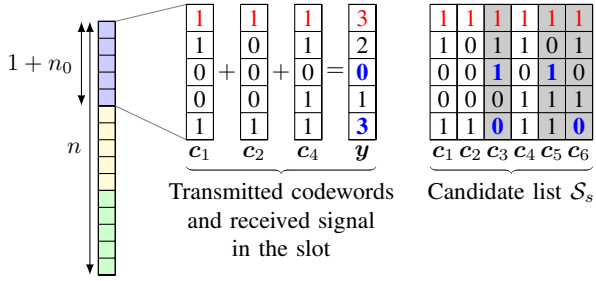
\begin{figure}
        \centering
        \begin{tikzpicture}[scale=.45,every node/.style={scale=0.9}]
            \def\R{1.6}
            \draw[latex-latex] (-1.25,3.5) -- node[left,midway] () {$n$} (-1.25,3.5 - 7.5);
            
            \draw[step=.5cm,draw=black,thin,fill=blue!20] (-1,3.5) grid (-.5,3.5-2.5) rectangle (-1,3.5);
            \draw[step=.5cm,black,thin,fill=yellow!20] (-1,3.5 - 2.5) grid (-.5,3.5 - 5) rectangle (-1,3.5 - 2.5);
            \draw[step=.5cm,black,thin,fill=green!20] (-1,3.5 - 5) grid (-.5,3.5 - 7.5) rectangle (-1,3.5 - 5);
            
            \draw[latex-latex] (-1.5,3.5) -- node[left,midway] () {$1+n_0$} (-1.5,3.5 - 2.5);

            \draw[color=gray] (-.5,3.5) -- (\R,2.5*\R);
            \draw[color=gray] (-.5,3.5 - 2.5) -- (\R,0);
				    
            \draw[step=.5cm*\R,draw=black,thin] (\R,0) grid (1.5*\R,2.5*\R) rectangle (\R,0);
            \node at (1.25*\R,-.25*\R) {$\cv_1$}; 
            \node at (1.25*\R,.25*\R) {1}; 
            \node at (1.25*\R,.75*\R) {0}; 
            \node at (1.25*\R,1.25*\R) {0}; 
            \node at (1.25*\R,1.75*\R) {1}; 
            \node at (1.25*\R,2.25*\R) {\red{1}}; 

            \node at (1.75*\R,1.25*\R) {$+$};
            
            \draw[step=.5cm*\R,draw=black,thin] (2*\R,0) grid (2.5*\R,2.5*\R) rectangle (2*\R,0);
            \node at (2.25*\R,-.25*\R) {$\cv_2$};
            \node at (2.25*\R,.25*\R) {1}; 
            \node at (2.25*\R,.75*\R) {0}; 
            \node at (2.25*\R,1.25*\R) {0}; 
            \node at (2.25*\R,1.75*\R) {0}; 
            \node at (2.25*\R,2.25*\R) {\red{1}}; 
            
            \node at (2.75*\R,1.25*\R) {$+$};
            
            \draw[step=.5cm*\R,draw=black,thin] (3*\R,0) grid (3.5*\R,2.5*\R) rectangle (3*\R,0);
            \node at (3.25*\R,-.25*\R) {$\cv_4$};
            \node at (3.25*\R,.25*\R) {1}; 
            \node at (3.25*\R,.75*\R) {1}; 
            \node at (3.25*\R,1.25*\R) {0}; 
            \node at (3.25*\R,1.75*\R) {1}; 
            \node at (3.25*\R,2.25*\R) {\red{1}}; 
            
            \node at (3.75*\R,1.25*\R) {$=$};
            
            \draw[step=.5cm*\R,draw=black,thin] (4*\R,0) grid (4.5*\R,2.5*\R) rectangle (4*\R,0);
            \node at (4.25*\R,-.25*\R) {$\yv$};
            \node at (4.25*\R,.25*\R) {\blue{\textbf{3}}}; 
            \node at (4.25*\R,.75*\R) {1}; 
            \node at (4.25*\R,1.25*\R) {\blue{\textbf{0}}}; 
            \node at (4.25*\R,1.75*\R) {2}; 
            \node at (4.25*\R,2.25*\R) {\red{3}}; 

            \draw [decorate, decoration = {calligraphic brace}] (8.5*\R,-0.5*\R) -- node[below=.1cm,midway] {Candidate list $\Sc_s$} (5.5*\R,-0.5*\R);

            \draw [decorate, decoration = {calligraphic brace}] (4.5*\R,-0.5*\R) -- node[below=.1cm,midway,align=center] {Transmitted codewords \\ and received signal \\ in the slot} (\R,-0.5*\R);
            
            \draw[step=.5cm*\R,draw=black,thin] (5.5*\R,0) grid (6*\R,2.5*\R) rectangle (5.5*\R,0);
            \node at (5.75*\R,-.25*\R) {$\cv_1$};
            \node at (5.75*\R,.25*\R) {1}; 
            \node at (5.75*\R,.75*\R) {0}; 
            \node at (5.75*\R,1.25*\R) {0}; 
            \node at (5.75*\R,1.75*\R) {1}; 
            \node at (5.75*\R,2.25*\R) {\red{1}}; 
            
            \draw[step=.5cm*\R,draw=black,thin] (6*\R,0) grid (6.5*\R,2.5*\R) rectangle (6*\R,0);
            \node at (6.25*\R,-.25*\R) {$\cv_2$};
            \node at (6.25*\R,.25*\R) {1}; 
            \node at (6.25*\R,.75*\R) {0}; 
            \node at (6.25*\R,1.25*\R) {0}; 
            \node at (6.25*\R,1.75*\R) {0}; 
            \node at (6.25*\R,2.25*\R) {\red{1}}; 

            \draw[step=.5cm*\R,draw=black,thin,fill=gray!40] (6.5*\R,0) grid (7*\R,2.5*\R) rectangle (6.5*\R,0);
            \node at (6.75*\R,-.25*\R) {$\cv_3$};
            \node at (6.75*\R,.25*\R) {\blue{\textbf{0}}}; 
            \node at (6.75*\R,.75*\R) {0}; 
            \node at (6.75*\R,1.25*\R) {\blue{\textbf{1}}}; 
            \node at (6.75*\R,1.75*\R) {1}; 
            \node at (6.75*\R,2.25*\R) {\red{1}}; 
           
            \draw[step=.5cm*\R,draw=black,thin] (7*\R,0) grid (7.5*\R,2.5*\R) rectangle (7*\R,0);
            \node at (7.25*\R,-.25*\R) {$\cv_4$};
            \node at (7.25*\R,.25*\R) {1}; 
            \node at (7.25*\R,.75*\R) {1}; 
            \node at (7.25*\R,1.25*\R) {0}; 
            \node at (7.25*\R,1.75*\R) {1}; 
            \node at (7.25*\R,2.25*\R) {\red{1}}; 

            \draw[step=.5cm*\R,draw=black,thin,fill=gray!40] (7.5*\R,0) grid (8.5*\R,2.5*\R) rectangle (7.5*\R,0);
            \node at (7.75*\R,-.25*\R) {$\cv_5$};
            \node at (7.75*\R,.25*\R) {1}; 
            \node at (7.75*\R,.75*\R) {1}; 
            \node at (7.75*\R,1.25*\R) {\blue{\textbf{1}}}; 
            \node at (7.75*\R,1.75*\R) {0}; 
            \node at (7.75*\R,2.25*\R) {\red{1}}; 
            \node at (8.25*\R,-.25*\R) {$\cv_6$};
            \node at (8.25*\R,.25*\R) {\blue{\textbf{0}}}; 
            \node at (8.25*\R,.75*\R) {1}; 
            \node at (8.25*\R,1.25*\R) {0}; 
            \node at (8.25*\R,1.75*\R) {1}; 
            \node at (8.25*\R,2.25*\R) {\red{1}}; 
        \end{tikzpicture}
        \caption{\revise{An example of the transmission and decoding in a slot~$s$ with $n_0 = 4$, $\rv{K}_s = 3$, and $\Sc_s = \{\cv_1,\cv_2,\cv_3,\cv_4,\cv_5,\cv_6\}$. The decoder reads $\rv{K}_s$ from the first (top) entry of the received signal $\yv$. It discards the codewords $\cv_3, \cv_5$, and $\cv_6$ from the candidate list $\Sc_s$ since they are incompatible with the third and/or the fifth entries of $\yv$. Only $\rv{K}_s = 3$ codewords remain in $\Sc_s$. Thus, the slot is resolved.}}
        \label{fig:example}
    \end{figure}

	If $\rv{K}_s > 0$, the receiver goes through the remaining channel uses with index~$j \in [2:n_0+1]$ of the slot and progressively discards incompatible codewords from the candidate list $\Sc_s$. Let $y(j)$ be the received signal in the $j$th channel use of the slot and $c(j)$ be the $j$th entry of codeword $\cv$. For each channel use $j$, if $y(j)=\rv{K}_s$, the decoder discards from $\Sc_s$ all codewords $\cv$ with $c(j) = 0$; similarly, if $y(j)= 0$, the decoder discards from $\Sc_s$ all codewords $\cv$ with $c(j) = 1$. We use $\log_2 K$ bits of each codeword to represent the user's signature,\footnote{In the unsourced model, the user identification task is deferred to upper layer protocols that spend at least $\log_2 K$ bits in the overhead of each packet.} so that different users transmit different codewords. Therefore, after the discarding process, if $\rv{K}_s$ codewords survive, they must be the transmitted codewords and the slot is resolved. \revise{An illustration of this per-slot decoder is given in Fig.~\ref{fig:example}.} Its \gls{MPR} capability is analyzed in the following theorem.
	%
   
	\begin{theorem}[\gls{MPR} capability] \label{th:MPR_expurgation}
		The proposed per-slot decoder can resolve a degree-$U$ slot with probability
		\begin{align}
			\pi_U = \revise{\EE\bigg[\Big(1-\frac{\Lambda'(1)}{N} \nu^{\rv{A}_U} (1-\nu)^{\rv{A}_0}\Big)^{M-U} \bigg]}\label{eq:piU}
		\end{align}
		where $\rv{A}_0$ and $\rv{A}_U$ are random nonnegative integers with joint \gls{pmf} given by 
		\begin{align}
			&\P[\rv{A}_0 = A_0, \rv{A}_U = A_U] = \frac{n_0!}{A_0!A_U!(n_0-A_0-A_U)!} \notag \\
			&\qquad  \cdot (1-\nu)^{U A_0} \nu^{U A_U} (1-(1-\nu)^U- \nu^U)^{n_0-A_0-A_U} \label{eq:pmf_A0AU}
		\end{align} 
		for $A_0 + A_U \le n_0$. In particular, if $\nu = 0.5$, it holds that
		\begin{align}
			\pi_U &= \revise{\EE\bigg[\Big(1-\frac{\Lambda'(1)}{N} 2^{-\rv{A}}\Big)^{M-U}\bigg]} \label{eq:piU_nu0.5} \\
            &\ge \pi_U' \triangleq \Big(1-\frac{\Lambda'(1)}{N} \big(1-2^{-U}\big)^{n_0}\Big)^{M-U}. \label{eq:piU_nu0.5_lb}
		\end{align}
        \revise{with $\rv{A} \sim \Bino(n_0,2^{1-U})$.}
	\end{theorem}
	\begin{IEEEproof}
        See~Appendix~\ref{app:MPR_expurgation}.
	\end{IEEEproof}

	\begin{remark} \label{rem:trade-off_nu}
        Due to symmetry, assume that $\nu \le 0.5$. 
        Compared to $\nu = 0.5$, setting $\nu < 0.5$ increases the chance that all active users transmit~$0$ and thus creates more discarding instances. This increases~$\pi_U$ for large~$U$. However, setting $\nu < 0.5$ also increases the probability $1-\nu$ that a nontransmitted codeword is compatible with the channel output in a channel use where the channel output is~$0$. This reduces~$\pi_U$ if~$U$ is small.  We shall illustrate this trade-off in Section~\ref{sec:numerical}.
	\end{remark}

	We are interested in the asymptotic regime where $M$, $K$, $N$, and $n_0$ all go to infinity and satisfy: i) $\mu K/N = G$ where $G$ is the channel load as in conventional analyses of \gls{IRSA}, ii) $n_0 = \beta \log_2M$ where $\beta^{-1} = \frac{\log_2M}{n_0} \le 1$ is the total number of bits transmitted per channel use in a slot, and iii) $M = D K^{1/\delta}$, $\delta \in (0,1)$, so that the per-user payload $\log_2(M/K)$ is split into a fixed amount of $\log_2 D$ bits and an increasing amount of $(1/\delta - 1) \log_2 K$ bits as~$K$ grows. We assume that the per-user payload increases with~$K$ because, if the users transmit a fixed number of bits over an infinite number of channel uses $n_0$, they achieve zero \gls{plr} but also zero information rate $\log_2(M/K)/n_0$. We refer to the defined regime as the $(G,D,\beta,\delta)$-asymptotic regime. In this regime, we characterize the \gls{MPR} capability of the per-slot decoder and the resulting decoding threshold for $\nu = 0.5$ as follows. 
	\begin{theorem}[Asymptotic \revise{lower bound on} \gls{MPR} capability and decoding threshold] \label{th:MPR_asymptotic}
		Let $\nu = 0.5$. In the $(G,D,\beta,\delta)$-asymptotic regime, it holds that
		\begin{align}
			&\revise{\lim\limits_{N\to\infty}} \pi_U \revise{\ \ge\ } \bar{\pi}_U \notag \\
			&
			=  \begin{cases}
				0, &\text{if~} U > - \log_2(1 - 2^{-(1-\delta)/\beta}), \\
				\revise{\frac{(1-2^{-U})^{\beta \log_2D}}{\exp(\mu^{-1} D G \Lambda'(1))}}, &\text{if~} U = - \log_2(1 - 2^{-(1-\delta)/\beta}), \\
				1, &\text{if~} U < - \log_2(1 - 2^{-(1-\delta)/\beta}).
			\end{cases} \label{eq:piU_asymptotic}
		\end{align}
		Furthermore, the decoding threshold $G^*(\Lambda)$ is \revise{lower bounded} by the largest value of $G$ such that
    	\begin{equation}
    		p > \lambda\bigg( 1 - e^{-G \Lambda'(1)p} \sum_{u=0}^{{\bar{T}_{\text{ED-MPR} - 1}}} \frac{\bar{\pi}_{u+1}}{u!} (p G \Lambda'(1))^u \bigg), \forall p \in (0,1],
    		\label{eq:MPR_dec_thres}
    	\end{equation}
    	where 
    	\begin{equation}
    		\bar{T}_{\text{ED-MPR}}=  \big\lfloor-\log_2\big(1 - 2^{-(1-\delta)/\beta}\big)\big\rfloor \label{eq:T}
    	\end{equation}
    	is \revise{a lower bound on} the maximum number of resolvable users in a slot.
	\end{theorem}
	\begin{IEEEproof}
		See~Appendix~\ref{app:MPR_asymptotic}.
	\end{IEEEproof}
	
	\begin{remark} \label{rem:region_interest}
		The regime of interest corresponds to a small $\beta$, so that the users collectively transmit close to $1$ bit per channel use in a slot. Furthermore, aiming for massive IoT applications where a large number of devices transmit small data volumes~\cite{Ericsson2020_cellular_IoT}, we are interested in $\delta$ close to $1$, so that the per-user payload increases slowly with the number of users.
	\end{remark}

	\begin{remark} \label{rem:PVTC}
		In~\cite{Paolini2022IRSA_BAC}, Paolini, Valentini, Tralli, and Chiani consider a similar \gls{IRSA}-based scheme, which we shall refer to as PVTC. While our random-coding scheme uses an \gls{iid} binary codebook, \cite{Paolini2022IRSA_BAC}~employs the BPR codebook~\cite{David1993adderchannel}, which is constructed as the parity-check matrix of a BCH code. This construction applies to $D = 1$ and $\beta$ being a positive integer. If $\beta$ is not an integer, one can apply the PVTC scheme designed for $\lfloor \beta \rfloor$. This allows the per-slot decoder to resolve slots of degree up to $T_{\text{PVTC}} = \lfloor\beta\rfloor$, i.e., $\pi_U = \ind{U \le \lfloor\beta\rfloor}$. Currently, there is no efficient decoder that attempts to resolve slots of degree higher than $\lfloor\beta\rfloor$ for this scheme. The capability to do so has not been discussed in~\cite{David1993adderchannel,Paolini2022IRSA_BAC}. It is easy to show that 
		\begin{equation} \label{eq:compate_T}
			\begin{cases}
				\bar{T}_{\text{ED-MPR}} > T_{\text{PVTC}}, &\text{if~} \delta > 1+ \beta \log_2(1-2^{-\lfloor \beta \rfloor - 1}), \\ 
				\bar{T}_{\text{ED-MPR}} < T_{\text{PVTC}}, &\text{if~} \delta < 1+ \beta \log_2(1-2^{-\lfloor \beta \rfloor}), \\
				\bar{T}_{\text{ED-MPR}} = T_{\text{PVTC}}, &\text{otherwise}. 
			\end{cases}
		\end{equation}
        \revise{Therefore, if $\delta > 1+ \beta \log_2(1-2^{-\lfloor \beta \rfloor - 1})$, our per-slot decoder resolves more colliding users than the PVTC scheme. This corresponds to the regime of interest noted in Remark~\ref{rem:region_interest}, i.e., $\beta$ is small and $\delta$ is close to $1$, as we shall illustrate in Section~\ref{sec:numerical}.} 
	\end{remark}

	\subsection{Erasure Decoding on the Full Graph (ED-FG)}
	\label{sec:PD_FG}
	\revise{In the decoder introduced in Section~\ref{sec:PD_MPR}, while decoding in a slot $s$, the candidate lists of all other slots remain unchanged.} 
    Notice that if a codeword $\cv$ is incompatible with at least one of the slots in $h(\cv)$, it cannot have been transmitted, and thus can be removed from $\Sc_s, \forall s \in h(\cv)$. Based on this observation, we augment the per-slot discarding decoder as follows. For each slot, after the discarding process, the decoder removes each discarded codeword from the candidate lists for all associated slots.  In this way, we shrink the candidate list in a slot, and thus increase the probability of resolving the slot in later iterations. 
	
	Consider a \emph{full} graph that contains all $M$ codewords as VNs. This is identical to the original graph if all codewords have been transmitted. The decoder has knowledge of this full graph but initially cannot distinguish between VNs corresponding to transmitted and nontransmitted codewords. The aforementioned improved decoder can be equivalently interpreted as erasure decoding on this full graph. Specifically, from the CN perspective, an edge is revealed if the corresponding codeword is either resolved (i.e., we know it has been transmitted) or it is incompatible with the channel output (i.e., we know it has not been transmitted).  From the VN perspective, if an edge has been revealed, we know whether the corresponding codeword has been transmitted. Thus all other edges are also revealed.

	\section{Information Rate} \label{sec:rate}
	As $\log_2 K$ bits are allocated for the user's signature, the information rate of each user is $R = \frac{\log_2(M/K)}{n}$ bits per channel use. The average sum rate is thus given by $R\sub{sum} = \E[\rv{K}\sub{a}] R = \frac{\mu K \log_2 (M/K)}{n}$. It follows from the frame structure that
	\begin{align}
		R\sub{sum} =  \frac{\mu K}{N} \cdot \frac{\log_2 (M/K)}{1+n_0}.
	\end{align} 
	The next theorem characterizes the asymptotic average sum rate achievable with vanishing \gls{plr}.
	\begin{theorem}[Achievable average sum rate]
		In the $(G,D,\beta,\delta)$-asymptotic regime, an IRSA-based scheme with degree distribution $\Lambda(x)$ and decoding threshold $G^*(\Lambda)$ can achieve the following average sum rate with vanishing \gls{plr}
		\begin{align}
			\lim\limits_{K\to\infty} {R}_{\rm sum} = \frac{1-\delta}{\beta} G, \label{eq:Rsum}
		\end{align}
		for every $D > 0$, $\beta \ge 1$, $\delta \in (0,1)$, and $G \le G^*(\Lambda)$.
	\end{theorem}
	\begin{IEEEproof}
		In the considered asymptotic regime, the average sum rate is
		$
			R_{\rm sum} \!=\! G\cdot \frac{\log_2(M/K)}{1+n_0} \!=\! G\cdot \frac{(1/\delta\!-\!1)\log_2 K + \log_2 D}{1+\frac{\beta}{\delta} \log_2 K } \to \frac{1-\delta}{\beta} G.
		$
		By definition of the decoding threshold $G^*(\Lambda)$, for any channel load $G\le G^*(\Lambda)$, the \gls{plr} vanishes.
	\end{IEEEproof}
	
	\section{Numerical Results} \label{sec:numerical}
\begin{figure}[t!]
		\centering
		\captionsetup[subfigure]{oneside}
		\subfloat[\centering $\Lambda(x) = x^2$]{\label{fig:plr_a}
			\begin{tikzpicture}[scale=.85]
				\begin{axis}[%
					width=1.7in,
					height=1.7in,
					scale only axis,
					unbounded coords=jump,
					xmin=0,
					xmax=3,
					xlabel style={font=\color{white!15!black},yshift=4pt},
					xlabel={Channel load $G = \mu K/N$},
					ymin=1e-6,
					ymax=1,
					ymode=log,
					ytick={1e0,1e-2,1e-4,1e-6},
					yticklabels={,,},
					xmajorgrids,
					ymajorgrids,
					legend style={at={(.01,0.99)}, anchor=north west, legend cell align=left, align=left, draw=white!15!black}
					]
					
					\addplot [line width = 1.5,color=blue]
					table[row sep=crcr]{%
                       0.200000000000000   0.000001800168316 \\
                       0.300000000000000   0.000004000032534 \\
                       0.400000000000000   0.000007738375404 \\
                       0.500000000000000   0.000012969299009 \\
                       0.600000000000000   0.000018899419788 \\
                       0.700000000000000   0.000026769174607 \\
                       0.800000000000000   0.000034839213499 \\
                       0.900000000000000   0.000046875857705 \\
                       1.000000000000000   0.000060125018635 \\
                       1.100000000000000   0.000076774906664 \\
                       1.200000000000000   0.000115451161576 \\
                       1.300000000000000   0.000394654226388 \\
						1.400000000000000   0.020803422201738 \\
						1.500000000000000   0.095202584464216 \\
						1.600000000000000   0.248246357959824 \\
						1.700000000000000   0.422277951386004 \\
						1.800000000000000   0.561281144303257 \\
						1.900000000000000   0.659523179972655 \\
						2.000000000000000   0.730263594719806 \\
						2.100000000000000   0.782868431645707 \\
						2.200000000000000   0.823951983950411 \\
						2.300000000000000   0.856298008866537 \\
						2.400000000000000   0.882108071820076 \\
						2.500000000000000   0.903023210090949 \\
						2.6000   9.2022e-01 \\ 
						2.7000   9.3387e-01 \\ 
						2.8000   9.4542e-01 \\ 
						2.9000   9.5464e-01 \\ 
						3.0000   9.6203e-01 \\ 
					};
					\node[rotate=80,color=blue] at (axis cs:1.23,1e-2) () {\small PVTC scheme};
					
					\addplot [line width = 1.5,color=black]
					table[row sep=crcr]{%
						0.100000000000000   0.001224918309603 \\
						0.200000000000000   0.003292975327554 \\
						0.300000000000000   0.007198405938702 \\
						0.400000000000000   0.016778169625599 \\
						0.500000000000000   0.047724642190666 \\
						0.600000000000000   0.130932387069082 \\
						0.700000000000000   0.266998964161827 \\
						0.800000000000000   0.410847744367482 \\
						0.900000000000000   0.534881679696336 \\
						1.000000000000000   0.633851111075780 \\
						1.100000000000000   0.711136017156416 \\
						1.200000000000000   0.771291383718876 \\
						1.300000000000000   0.818565798570239 \\
						1.400000000000000   0.855252392863666 \\
						1.500000000000000   0.884330980744279 \\
						1.600000000000000   0.907503714895920 \\
						1.700000000000000   0.925626366899658 \\
						1.800000000000000   0.939989484133708 \\
						1.900000000000000   0.951506890616654 \\
						2.000000000000000   0.960796792611231 \\
						2.100000000000000   0.968288317659945 \\
						2.200000000000000   0.974219521752913 \\
						2.300000000000000   0.979054958033598 \\
						2.400000000000000   0.982944504825572 \\
						2.500000000000000   0.986152530533279 \\
						2.600000000000000   0.988713975143885 \\
						2.700000000000000   0.990783982824771 \\
						2.800000000000000   0.992482331198644 \\
						2.900000000000000   0.993874445553481  \\
						3.000000000000000   0.995006326912346 \\
					};
					\node[rotate=65,color=black] at (axis cs:.38,6e-2) () {\small Original IRSA};
					
					\addplot [line width = 1.5,color=red]
					table[row sep=crcr]{%
					0.600000000000000   0.000001479995752 \\
                      0.700000000000000   0.000002694228989 \\
                      0.800000000000000   0.000004629927409 \\
						0.900000000000000   0.000006406741261 \\
                      1.000000000000000   0.000008217085154 \\
                      1.100000000000000   0.000012447027385 \\
                      1.200000000000000   0.000017377243106 \\
                      1.300000000000000   0.000023120079816 \\
                      1.400000000000000   0.000026161152982 \\
                      1.500000000000000   0.000035598919816 \\
                      1.600000000000000   0.000048072186960 \\
                      1.700000000000000   0.000067083703583 \\
                      1.800000000000000   0.000087052324763 \\
                      1.900000000000000   0.000580183618292 \\
                      2.000000000000000   0.009583500125798 \\
                      2.100000000000000   0.068904325781189 \\
                      2.200000000000000   0.221303030108341 \\
                      2.300000000000000   0.410888487425292 \\
                      2.400000000000000   0.568833210243705 \\
                      2.500000000000000   0.631887134565282 \\
                      2.600000000000000   0.720995760918903 \\
                       2.700000000000000   0.793996675514684 \\
                       2.800000000000000   0.842280149877071 \\
                       2.900000000000000   0.882045442782465 \\
                       3.000000000000000   0.907945221947286 \\
					};
					\node[rotate=77,color=red] at (axis cs:1.85,6e-3) () {\small Our scheme, ED-MPR};
					
					\addplot [line width = 1.5,color=OliveGreen]
					table[row sep=crcr]{%
						1.6000   0.000002987786435 \\ 
						1.8000   3.4430e-06 \\ 
						1.9000   4.3181e-06 \\ 
						2.0000   4.4926e-06 \\ 
						2.1000   6.4012e-06 \\ 
						2.2000   5.2550e-05 \\ 
						2.3000   0.001095891791346 \\ 
						2.4000   0.013378354837907 \\ 
						2.5000   0.052206416060324 \\ 
						2.6000   0.206189137047829 \\ 
						2.7000   0.458472275984024 \\ 
						2.8000   0.667985825595000 \\ 
						2.9000   0.782249312687789 \\ 
						3.0000   0.840895020273581 \\ 
					};
					\node[rotate=80,color=OliveGreen] at (axis cs:2.45,6e-4) () {\small Our scheme, ED-FG};
					
					
					
				\end{axis}
			\end{tikzpicture}
		}
		\hspace{-.5cm}
		\captionsetup[subfigure]{oneside}
		\subfloat[\centering $\Lambda(x) = x^3$]{
			\label{fig:plr_b}
			\begin{tikzpicture}[scale=.85]
				\begin{axis}[%
					width=1.7in,
					height=1.7in,
					scale only axis,
					unbounded coords=jump,
					xmin=0,
					xmax=3,
					xlabel style={font=\color{white!15!black},yshift=4pt},
					xlabel={Channel load $G = \mu K/N$},
					ymin=1e-6,
					ymax=1,
					ymode=log,
					yminorticks=true,
					ytick={1e0,1e-2,1e-4,1e-6},
					ylabel style={font=\color{white!15!black},yshift=-18pt},
					ylabel={PLR},
					axis background/.style={fill=white},
					title style={font=\bfseries},
					xmajorgrids,
					ymajorgrids,
					legend style={at={(.01,0.99)}, anchor=north west, legend cell align=left, align=left, draw=white!15!black}
					]
					
					\addplot [line width = 1.5,color=black]
					table[row sep=crcr]{%
						0.100000000000000   0.000015402780972 \\
						0.200000000000000   0.000031623282065 \\
						0.300000000000000   0.000050566361583 \\
						0.400000000000000   0.000070063033355 \\
						0.500000000000000   0.000123815789061 \\
						0.600000000000000   0.004347571068982 \\
						0.700000000000000   0.094316670097961 \\
						0.800000000000000   0.388386821381553 \\
						0.900000000000000   0.646566411977005 \\
						1.000000000000000   0.779028815859255 \\
						1.100000000000000   0.852667135251769 \\
						1.200000000000000   0.897832260035415 \\
						1.300000000000000   0.928297108775040 \\
						1.400000000000000   0.948970818721066 \\
						1.500000000000000   0.963218067102386 \\
						1.600000000000000   0.973399445066187 \\
						1.700000000000000   0.980644758392464 \\
						1.800000000000000   0.985861329242355 \\
						1.900000000000000   0.989651435615898 \\
						2.000000000000000   0.992365681393433 \\
						2.100000000000000   0.994390742159594 \\
						2.200000000000000   0.995856420155059 \\
						2.300000000000000   0.996951663919501 \\
						2.400000000000000   0.997748039944295 \\
						2.500000000000000   0.998339752748028 \\
						2.600000000000000   0.998771399179175 \\
						2.700000000000000   0.999093791488377 \\
						2.800000000000000   0.999326949180869 \\
						2.900000000000000   0.999504464823820 \\
						3.000000000000000   0.999635746976700 \\
					};
					\node[rotate=83,color=black] at (axis cs:.46,5e-3) () {\small Original IRSA};
					
				 	\addplot [line width = 1.5,color=blue]
				 	table[row sep=crcr]{%
				 	    1.100000000000000   0.000002677157549 \\
                       1.200000000000000   0.000326185773288 \\
                       1.300000000000000   0.013537413502311 \\
                       1.400000000000000   0.136400879496418 \\
                       1.500000000000000   0.424134079404779 \\
                       1.600000000000000   0.692394003868170 \\
                       1.700000000000000   0.817014567060881 \\
                       1.800000000000000   0.877108002830108 \\
                       1.900000000000000   0.913217195058117 \\
                       2.000000000000000   0.936098603413345 \\
                       2.100000000000000   0.953021494670050 \\
                       2.200000000000000   0.965215141133938 \\
                       2.300000000000000   0.973755712394967 \\
                       2.400000000000000   0.980336813493545 \\
                       2.500000000000000   0.984999234562935 \\
                       2.600000000000000   0.988635869391595 \\
                       2.700000000000000   0.991465468006150 \\
                       2.800000000000000   0.993498267556455 \\
                       2.900000000000000   0.995138414352865 \\
                       3.000000000000000   0.996264695028951 \\
					};
					\node[rotate=82.7,color=blue] at (axis cs:1.1,2e-3) () {\small PVTC scheme};
					
					\addplot [line width = 1.5,color=red]
					table[row sep=crcr]{%
1.0e+00 0.0000e+00 \\ 
1.1e+00 0.0000e+00 \\ 
1.2e+00 0.0000e+00 \\ 
1.3e+00 0.0000e+00 \\ 
1.4e+00 0.000001285961282 \\ 
1.5e+00 0.000026373068289 \\ 
1.6e+00 6.4022e-04 \\ 
1.7e+00 3.5893e-02 \\ 
1.8e+00 1.6416e-01 \\ 
1.9e+00 4.6291e-01 \\ 
2.0e+00 7.1324e-01 \\ 
2.1e+00 8.4378e-01 \\ 
2.2e+00 8.9046e-01 \\ 
2.3e+00 9.2292e-01 \\ 
2.4e+00 9.4179e-01 \\ 
2.5e+00 9.5383e-01 \\ 
2.6e+00 9.6538e-01 \\ 
2.7e+00 9.7467e-01 \\ 
2.8e+00 9.7967e-01 \\ 
2.9e+00 9.8525e-01 \\ 
3.0e+00 9.8867e-01 \\ 
					};
					\node[rotate=83,color=red] at (axis cs:1.45,5e-4) () {\small Our scheme, ED-MPR};
					
					\addplot [line width = 1.5,color=OliveGreen]
					table[row sep=crcr]{%
					    1.76 1.2188e-06 \\
					    1.8 3.7721e-06 \\
					    1.900000000000000   0.000475822492630 \\
					   2.000000000000000   0.007976744186047 \\
                       2.100000000000000   0.133844036045142 \\
                       2.200000000000000   0.364298644104535 \\
                       2.300000000000000   0.735275567734525 \\
                       2.400000000000000   0.886148639705405 \\
                       2.500000000000000   0.912101573242571 \\
                       2.600000000000000   0.945192833505606 \\
                       2.700000000000000   0.963801487890760 \\
                       2.800000000000000   0.973463387601126 \\
                       2.900000000000000   0.981196620226827 \\
                       3.000000000000000   0.984354356689450 \\
					};
					\node[rotate=82.5,color=OliveGreen] at (axis cs:2.07,5e-4) () {\small Our scheme, ED-FG};
					
					
					
				\end{axis}
			\end{tikzpicture}
		}
		\caption{The \gls{plr} achieved with our scheme, original IRSA, and the PVTC scheme for $\nu = 0.5$, $N = 200$, $\mu = 0.2$, $\mu K/N = G$, $M = K^{1/\delta}$, and $n_0 = \beta \log_2 M$ with $(\beta,\delta) = (2,0.9)$.}
		\label{fig:PLR}
	\end{figure}
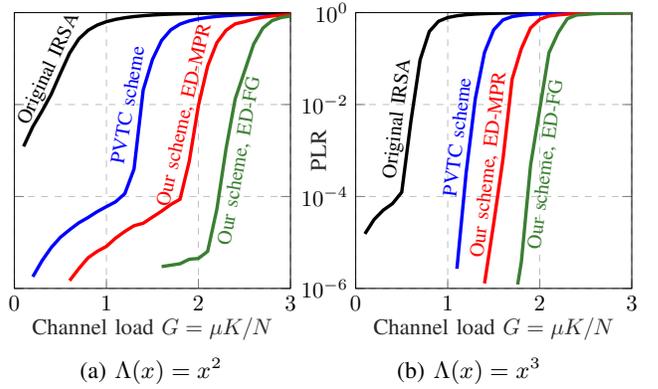

	We numerically evaluate the random-coding bound on the performance of our proposed decoders and compare it with the original \gls{IRSA} and the PVTC scheme~\cite{Paolini2022IRSA_BAC}. We consider $D=1$ to enable a comparison with the PVTC scheme. In Fig.~\ref{fig:PLR}, we plot the \gls{plr} (obtained via Monte-Carlo simulation) achieved with these schemes as a function of the channel load $G \!=\! \mu K/N$ for $\nu \!=\! 0.5$, $N \!=\! 200$, $\mu \!=\! 0.2$, and $\Lambda(x) \in \{x^2,x^3\}$. \revise{For our scheme, we implemented the per-slot discarding decoder.} We set $M = K^{1/\delta}$, $n_0 = \beta \log_2 M$ with $(\beta,\delta) = (2,0.9)$, and increase $K$. 
	We observe that our proposed scheme with ED-MPR~(Section~\ref{sec:PD_MPR}) significantly outperforms the original \gls{IRSA} and the PVTC scheme. The \gls{plr} can be further reduced notably with ED-FG~(Section~\ref{sec:PD_FG}). For example, to achieve a \gls{plr} of $10^{-4}$ with $\Lambda(x) = x^2$, our scheme with ED-MPR and ED-FG can support around $52\%$ and $91\%$ more users per slot in average, respectively, than the PVTC scheme.
	
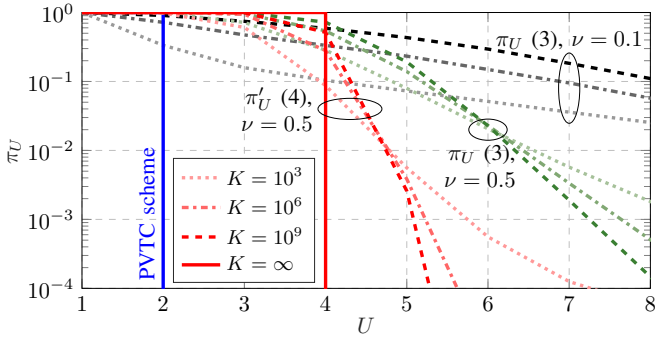
\begin{figure}[t!]
		\centering
		\begin{tikzpicture}[scale=.85]
			\begin{axis}[%
				width=3.5in,
				height=1.7in,
				scale only axis,
				unbounded coords=jump,
				xmin=1,
				xmax=8,
				xtick={1,2,3,4,5,6,7,8,9,10},
				xlabel style={font=\color{white!15!black},yshift=4pt},
				xlabel={$U$},
				ymin=1e-4,
				ymax=1,
				ymode=log,
				yminorticks=true,
				ylabel style={font=\color{white!15!black},yshift=-4pt},
				ylabel={$\pi_U$},
				axis background/.style={fill=white},
				title style={font=\bfseries},
				xmajorgrids,
				ymajorgrids,
				legend style={font=\small,at={(.41,0.02)}, anchor=south east, legend cell align=left, align=left, draw=white!15!black}
				]

    \addplot [line width = 1.5,dotted,color=black!40, forget plot]
				table[row sep=crcr]{%
                    1 1 \\ 
                    2 3.3430e-01 \\ 
                    3 1.5843e-01 \\ 
                    4 1.0384e-01 \\ 
                    5 7.2886e-02 \\ 
                    6 5.1374e-02 \\ 
                    7 3.5919e-02 \\ 
                    8 2.5391e-02 \\ 
                    9 1.8108e-02 \\ 
                    10 1.2950e-02 \\ 
				};
				
				\addplot [line width = 1.5,dashdotted,color=black!60,forget plot]
				table[row sep=crcr]{%
                    1 1 \\ 
                    2 7.2092e-01 \\ 
                    3 4.7305e-01 \\ 
                    4 3.3377e-01 \\ 
                    5 2.3294e-01 \\ 
                    6 1.4923e-01 \\ 
                    7 9.5114e-02 \\ 
                    8 5.7286e-02 \\ 
                    9 3.7103e-02 \\ 
                    10 1.9230e-02 \\ 
				};
				
				\addplot [line width = 1.5,dashed,color=black, forget plot]
				table[row sep=crcr]{%
                    1 1 \\ 
                    2 9.0614e-01 \\ 
                    3 7.4505e-01 \\ 
                    4 5.9051e-01 \\ 
                    5 4.3318e-01 \\ 
                    6 2.9490e-01 \\ 
                    7 1.8318e-01 \\ 
                    8 1.1020e-01 \\ 
                    9 5.6463e-02 \\ 
                    10 2.6760e-02 \\ 
				};

				\addplot [line width = 1.5,color=blue,forget plot]
				table[row sep=crcr]{%
					1 1 \\
					2 1  \\ 
					2 1e-30 \\
					3 1e-30 \\
					4 1e-30 \\
					5 1e-30 \\
					6 1e-30 \\
					7 1e-30 \\
					8 1e-30 \\
				};
				\node[rotate=90,color=blue] at (axis cs:1.8,1.2e-3) () {PVTC scheme};
				
				\addplot [line width = 1.5,dotted,color=OliveGreen!40,forget plot]
				table[row sep=crcr]{%
                    1 1.0000e+00 \\ 
                    2 9.8654e-01 \\ 
                    3 7.1766e-01 \\ 
                    4 2.8425e-01 \\ 
                    5 7.9424e-02 \\ 
                    6 2.0587e-02 \\ 
                    7 5.7212e-03 \\ 
                    8 1.8171e-03 \\ 
                    9 6.7253e-04 \\ 
                    10 2.8799e-04 \\ 
				};

				\addplot [line width = 1.5,dashdotted,color=OliveGreen!60,forget plot]
				table[row sep=crcr]{%
                    1 1.0000e+00 \\ 
                    2 9.9994e-01 \\ 
                    3 9.5544e-01 \\ 
                    4 5.4205e-01 \\ 
                    5 1.3908e-01 \\ 
                    6 2.2962e-02 \\ 
                    7 3.3448e-03 \\ 
                    8 5.0806e-04 \\ 
                    9 8.6775e-05 \\ 
                    10 1.7066e-05 \\ 
				};

				\addplot [line width = 1.5,dashed,OliveGreen=red,forget plot]
				table[row sep=crcr]{%
                    1 1.0000e+00 \\ 
                    2 1.0000e+00 \\ 
                    3 9.9417e-01 \\ 
                    4 7.2401e-01 \\ 
                    5 1.9250e-01 \\ 
                    6 2.2793e-02 \\ 
                    7 1.8988e-03 \\ 
                    8 1.4691e-04 \\ 
                    9 1.2095e-05 \\ 
                    10 1.1222e-06 \\ 
				};

                \addplot [line width = 1.5,dotted,color=red!40]
				table[row sep=crcr]{%
					1.000000000000000   0.999998716453755 \\
					2.000000000000000   0.985704883364277 \\
					3.000000000000000   0.607253848600309 \\
					4.000000000000000   0.087311764003257 \\
					5.000000000000000   0.005604043463218 \\
					6.000000000000000   0.000558060375332 \\
					7.000000000000000   0.000125440162275 \\
					8.000000000000000   0.000053772239777 \\
					9.000000000000000   0.000034309394787 \\
					10.000000000000000   0.000027272709009 \\
				};
				\addlegendentry{$K = 10^3$};

				\addplot [line width = 1.5,dashdotted,color=red!60]
				table[row sep=crcr]{%
					1.000000000000000   1.000000000000000 \\
					2.000000000000000   0.999944616522908 \\
					3.000000000000000   0.944578703818350 \\
					4.000000000000000   0.280388171343497 \\
					5.000000000000000   0.003844399510996 \\
					6.000000000000000   0.000010917248759 \\
					7.000000000000000   0.000000082844043 \\
					8.000000000000000   0.000000003536371 \\
					9.000000000000000   0.000000000588507 \\
					10.000000000000000   0.000000000226183 \\
				};
				\addlegendentry{$K = 10^6$};

				\addplot [line width = 1.5,dashed,color=red]
				table[row sep=crcr]{%
					1.000000000000000   1.000000000000000 \\
					2.000000000000000   1.000000000000000 \\
					3.000000000000000   0.993511901604033 \\
					4.000000000000000   0.515671274451722 \\
					5.000000000000000   0.002583147071888 \\
					6.000000000000000   0.000000027552138 \\
					7.000000000000000   0.000000000000145 \\
					8.000000000000000   0.000000000000000 \\
					9.000000000000000   0.000000000000000 \\
					10.000000000000000   0.000000000000000 \\
				};
				\addlegendentry{$K = 10^9$};
				
				                \addplot [line width = 1.5,color=red]
				table[row sep=crcr]{%
					1 1 \\
					2 1  \\ 
					3 1 \\
					4 1 \\
					4 1e-30 \\
					5 1e-30 \\
					6 1e-30 \\
					7 1e-30 \\
					8 1e-30 \\
				};
				\addlegendentry{$K = \infty$};

				\node[ellipse, draw, minimum width = 3mm, minimum height = 11mm] () at (axis cs:7,8e-2) {};
				\node[align=center] at (axis cs:7,4e-1) () {$\pi_U$~\eqref{eq:piU_nu0.5}, $\nu = 0.1$};
				
				\node[ellipse, draw, minimum width = 6mm, minimum height = 3mm] () at (axis cs:6,2e-2) {};
				\node[align=right] at (axis cs:5.9,6e-3) () {$\pi_U$~\eqref{eq:piU_nu0.5}, \\ $\nu = 0.5$};

                \node[ellipse, draw, minimum width = 10mm, minimum height = 3mm] () at (axis cs:4.3,4e-2) {};
				\node[align=right] at (axis cs:3.4,4e-2) () {$\pi'_U$~\eqref{eq:piU_nu0.5_lb}, \\ $\nu = 0.5$};
			\end{axis}
		\end{tikzpicture}
		\caption{The probability of resolving degree-$U$ slots, $\pi_U$, achieved by our proposed scheme with ED-MPR and the PVTC scheme for $\Lambda(x) = x^2$, $\nu \in \{0.1,0.5\}$, $K/N = \mu^{-1}G = 2.5$, $M = K^{1/\delta}$, and $n_0 = \beta \log_2 M$ with $(\beta,\delta) = (2,0.9)$.}
		\label{fig:pi_U}
	\end{figure}

	The advantage of our scheme over the PVTC scheme for the considered setting comes from the capability to resolve more colliding users. To illustrate this, we consider a similar setting as in Fig.~\subref*{fig:plr_a} with $(\beta, \delta,\Lambda(x))=(2,0.9,x^2)$ and plot in Fig.~\ref{fig:pi_U} the probability of resolving degree-$U$ slots, $\pi_U$, \revise{and its lower bound $\pi'_U$}, given in Theorem~\ref{th:MPR_expurgation}. We assume that $\mu^{-1} G = 2.5$, i.e., $N = K/2.5$. The PVTC scheme resolves slots of degree up to $\beta = 2$. For finite $K$, our scheme with ED-MPR resolves most of degree-$2$ slots, and can further resolve slots of higher degrees.  As $K$ becomes large, with $\nu = 0.5$, $\pi'_U$ converges to $\ind{U \le \bar{T}_{\text{ED-MPR}}}$ with $\bar{T}_{\text{ED-MPR}} = 4$ obtained using~\eqref{eq:T}. This agrees with Theorem~\ref{th:MPR_asymptotic}. 
    With $\nu = 0.1$, our scheme resolves high-degree slots with higher probability but low-degree slots with lower probability. This is due to the trade-off mentioned in Remark~\ref{rem:trade-off_nu}. For $\Lambda(x) \in \{x^2, x^3\}$, the majority of the slots have low degree, thus $\nu=0.5$ leads to lower \gls{plr} than $\nu \ne 0.5$. 

\begin{figure}[t!]
		\centering
		\begin{tikzpicture}[scale=.85]
			\begin{axis}[%
				width=3.4in,
				height=1.6in,
				scale only axis,
				unbounded coords=jump,
				xmin=1,
				xmax=10,
				xlabel style={font=\color{white!15!black}},
				xlabel={\small $\beta = \big(\frac{\log_2M}{n_0}\big)^{-1}$},
				ticklabel style = {font=\footnotesize},
				ymin=0,
				ymax=1,
				yminorticks=true,
				ylabel style={font=\color{white!15!black}},
				ylabel={\small $\delta = \frac{\ln K}{\ln M}$},
				axis background/.style={fill=white},
				title style={font=\bfseries},
				xmajorgrids,
				ymajorgrids,
				legend style={at={(.99,0.02)}, anchor=south east, legend cell align=left, align=left, draw=white!15!black}
				]
				
				\addplot [name path = A, line width = 1,color=blue]
				table[row sep=crcr]{%
					1.000000000000000                   0 \\
					1.990990990990991  -0.990990990990991 \\
					2.000000000000000   0.169925001442312 \\
					2.990990990990991  -0.241373421266452 \\
					3.000000000000000   0.422064766172812 \\
					3.990990990990991   0.231155229473141 \\
					4.000000000000000   0.627562382434074 \\
					4.990990990990991   0.535291801505579 \\
					5.000000000000000   0.770981551934376 \\
					5.990990990990991   0.725590508173622 \\
					6.000000000000000   0.863679540999499 \\
					6.990990990990991   0.841164149873290 \\
					7.000000000000000   0.920792807405161 \\
					7.990990990990991   0.909579433936136 \\
					8.000000000000000   0.954827494870863 \\
					8.990990990990991   0.949231801667930 \\
					9.000000000000000   0.974615328438592 \\
					9.990990990990991   0.971820219457856 \\
					10.000000000000000   0.985904297453286 \\
				};
				
				\addplot [name path = B, line width = 1,color=red]
				table[row sep=crcr]{%
					1.000000000000000   0.584962500721156 \\
					1.990990990990991   0.173664078012392 \\
					2.000000000000000   0.614709844115208 \\
					2.990990990990991   0.423800307415537 \\
					3.000000000000000   0.720671786825556 \\
					3.990990990990991   0.628401205897060 \\
					4.000000000000000   0.816785241547501 \\
					4.990990990990991   0.771394197786747 \\
					5.000000000000000   0.886399617499582 \\
					5.990990990990991   0.863884226373373 \\
					6.000000000000000   0.932108120632995 \\
					6.990990990990991   0.920894747163970 \\
					7.000000000000000   0.960474058012006 \\
					7.990990990990991   0.954878364809072 \\
					8.000000000000000   0.977435847500971 \\
					8.990990990990991   0.974640738520235 \\
					9.000000000000000   0.987313867707958 \\
					9.990990990990991   0.985916996284410 \\
					10.000000000000000   0.992953870234106 \\
				};
				
				\addplot [name path = X]
				table[row sep=crcr]{%
					1 0 \\
					10 0 \\
				};
				
				\addplot [name path = Y]
				table[row sep=crcr]{%
					1 1 \\
					10 1 \\
				};
				
				\addplot [green!30, fill opacity=0.3] fill between [of = A and B];
				\addplot [blue!30, fill opacity=0.3] fill between [of = A and X];
				\addplot [red!30, fill opacity=0.3] fill between [of = B and Y];
				
				\node[] at (axis cs:7,.43) () {$\bar{T}_{\text{ED-MPR}} < T_{\text{PVTC}}$};
				\node[] at (axis cs:3.2,.9) () {$\bar{T}_{\text{ED-MPR}} > T_{\text{PVTC}}$};
			\end{axis}
		\end{tikzpicture}
		\caption{The $(\beta,\delta)$ regions where either $\bar{T}_{\text{ED-MPR}}$ or $T_{\text{PVTC}}$ is larger for $D = 1$ and $\nu = 0.5$. In the middle (green) region, 
        $\bar{T}_{\text{ED-MPR}} = T_{\text{PVTC}}$. \revise{When $\bar{T}_{\text{ED-MPR}} > T_{\text{PVTC}}$ (depicted in red), our scheme with ED-MPR guarantees to resolve more colliding users than the PVTC scheme} in the $(G,D,\beta,\delta)$-asymptotic regime.}
		\label{fig:T_region}
	\end{figure}
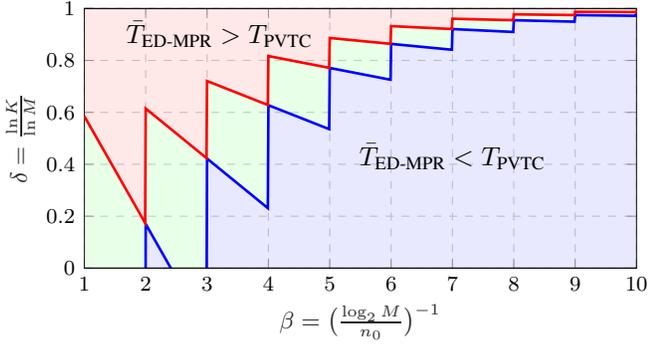

	Hereafter, we focus on the $(G,D,\beta,\delta)$-asymptotic. In this regime, the PVTC scheme can resolve slots of degree up to $T_{\text{PVTC}} = \lfloor \beta \rfloor$, \revise{while our scheme with ED-MPR guarantees to resolve slots of degree up to $\bar{T}_{\text{ED-MPR}}$ given in~\eqref{eq:T}}. In Fig.~\ref{fig:T_region}, we depict the regions of $(\beta,\delta)$ where either $T_{\text{PVTC}}$ or $\bar{T}_{\text{ED-MPR}}$ is larger, or they are equal, as characterized in~\eqref{eq:compate_T}, for $\nu = 0.5$. We see that our scheme with ED-MPR resolves more colliding users than the PVTC scheme if $\beta$ is small and $\delta$ is close to $1$. (We have seen in Fig.~\ref{fig:pi_U} that this is the case for $(\beta,\delta) = (2,0.9)$.) As noted in Remark~\ref{rem:region_interest}, this is the regime of interest.
	
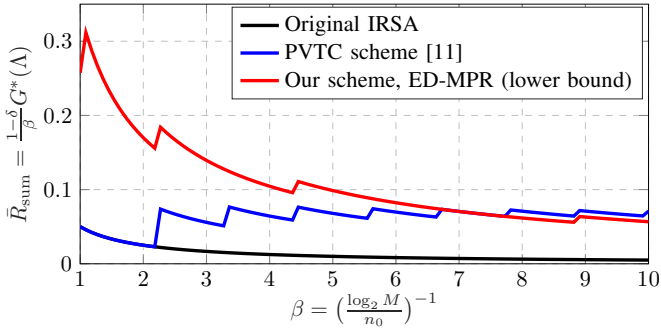
\begin{figure}[t!]
		\centering
		\begin{tikzpicture}[scale=.85]
			
			\begin{axis}[%
				width=3.5in,
				height=1.6in,
				scale only axis,
				unbounded coords=jump,
				xmin=1,
				xmax=10,
				xlabel style={font=\color{white!15!black},yshift=4pt},
				xlabel={$\beta = \big(\frac{\log_2M}{n_0}\big)^{-1}$},
				ymin=0,
				ymax=.35,
				yminorticks=true,
				ylabel style={font=\color{white!15!black},yshift=-4pt},
				ylabel={$\bar{R}_{\rm sum} = \frac{1-\delta}{\beta}G^*(\Lambda)$},
				axis background/.style={fill=white},
				title style={font=\bfseries},
				xmajorgrids,
				ymajorgrids,
				legend style={at={(.99,0.99)}, anchor=north east, legend cell align=left, align=left, draw=white!15!black}
				]
				
				\addplot [line width = 1.5,color=black]
				table[row sep=crcr]{%
					1.0000   5.0046e-02 \\ 
					1.0909   4.5875e-02 \\ 
					1.1818   4.2347e-02 \\ 
					1.2727   3.9322e-02 \\ 
					1.3636   3.6700e-02 \\ 
					1.4545   3.4407e-02 \\ 
					1.5455   3.2383e-02 \\ 
					1.6364   3.0584e-02 \\ 
					1.7273   2.8974e-02 \\ 
					1.8182   2.7525e-02 \\ 
					1.9091   2.6215e-02 \\ 
					2.0000   2.5023e-02 \\ 
					2.0909   2.3935e-02 \\ 
					2.1818   2.2938e-02 \\ 
					2.2727   2.2020e-02 \\ 
					2.3636   2.1173e-02 \\ 
					2.4545   2.0389e-02 \\ 
					2.5455   1.9661e-02 \\ 
					2.6364   1.8983e-02 \\ 
					2.7273   1.8350e-02 \\ 
					2.8182   1.7758e-02 \\ 
					2.9091   1.7203e-02 \\ 
					3.0000   1.6682e-02 \\ 
					3.0909   1.6191e-02 \\ 
					3.1818   1.5729e-02 \\ 
					3.2727   1.5292e-02 \\ 
					3.3636   1.4879e-02 \\ 
					3.4545   1.4487e-02 \\ 
					3.5455   1.4116e-02 \\ 
					3.6364   1.3763e-02 \\ 
					3.7273   1.3427e-02 \\ 
					3.8182   1.3107e-02 \\ 
					3.9091   1.2802e-02 \\ 
					4.0000   1.2511e-02 \\ 
					4.0909   1.2233e-02 \\ 
					4.1818   1.1968e-02 \\ 
					4.2727   1.1713e-02 \\ 
					4.3636   1.1469e-02 \\ 
					4.4545   1.1235e-02 \\ 
					4.5455   1.1010e-02 \\ 
					4.6364   1.0794e-02 \\ 
					4.7273   1.0587e-02 \\ 
					4.8182   1.0387e-02 \\ 
					4.9091   1.0195e-02 \\ 
					5.0000   1.0009e-02 \\ 
					5.0909   9.8304e-03 \\ 
					5.1818   9.6580e-03 \\ 
					5.2727   9.4915e-03 \\ 
					5.3636   9.3306e-03 \\ 
					5.4545   9.1751e-03 \\ 
					5.5455   9.0247e-03 \\ 
					5.6364   8.8791e-03 \\ 
					5.7273   8.7382e-03 \\ 
					5.8182   8.6016e-03 \\ 
					5.9091   8.4693e-03 \\ 
					6.0000   8.3410e-03 \\ 
					6.0909   8.2165e-03 \\ 
					6.1818   8.0957e-03 \\ 
					6.2727   7.9783e-03 \\ 
					6.3636   7.8644e-03 \\ 
					6.4545   7.7536e-03 \\ 
					6.5455   7.6459e-03 \\ 
					6.6364   7.5412e-03 \\ 
					6.7273   7.4393e-03 \\ 
					6.8182   7.3401e-03 \\ 
					6.9091   7.2435e-03 \\ 
					7.0000   7.1494e-03 \\ 
					7.0909   7.0578e-03 \\ 
					7.1818   6.9684e-03 \\ 
					7.2727   6.8813e-03 \\ 
					7.3636   6.7964e-03 \\ 
					7.4545   6.7135e-03 \\ 
					7.5455   6.6326e-03 \\ 
					7.6364   6.5536e-03 \\ 
					7.7273   6.4765e-03 \\ 
					7.8182   6.4012e-03 \\ 
					7.9091   6.3276e-03 \\ 
					8.0000   6.2557e-03 \\ 
					8.0909   6.1855e-03 \\ 
					8.1818   6.1167e-03 \\ 
					8.2727   6.0495e-03 \\ 
					8.3636   5.9838e-03 \\ 
					8.4545   5.9194e-03 \\ 
					8.5455   5.8564e-03 \\ 
					8.6364   5.7948e-03 \\ 
					8.7273   5.7344e-03 \\ 
					8.8182   5.6753e-03 \\ 
					8.9091   5.6174e-03 \\ 
					9.0000   5.5607e-03 \\ 
					9.0909   5.5051e-03 \\ 
					9.1818   5.4505e-03 \\ 
					9.2727   5.3971e-03 \\ 
					9.3636   5.3447e-03 \\ 
					9.4545   5.2933e-03 \\ 
					9.5455   5.2429e-03 \\ 
					9.6364   5.1934e-03 \\ 
					9.7273   5.1449e-03 \\ 
					9.8182   5.0973e-03 \\ 
					9.9091   5.0505e-03 \\ 
					10.0000   5.0046e-03 \\ 
				};
				\addlegendentry{Original IRSA};
				
				\addplot [line width = 1.5,color=blue]
				table[row sep=crcr]{%
					1.0000   5.0028e-02 \\ 
					1.0909   4.5859e-02 \\ 
					1.1818   4.2332e-02 \\ 
					1.2727   3.9308e-02 \\ 
					1.3636   3.6687e-02 \\ 
					1.4545   3.4394e-02 \\ 
					1.5455   3.2371e-02 \\ 
					1.6364   3.0573e-02 \\ 
					1.7273   2.8964e-02 \\ 
					1.8182   2.7516e-02 \\ 
					1.9091   2.6205e-02 \\ 
					2.0000   2.5014e-02 \\ 
					2.0909   2.3927e-02 \\ 
					2.1818   2.2930e-02 \\ 
					2.2727   7.3720e-02 \\ 
					2.3636   7.0885e-02 \\ 
					2.4545   6.8259e-02 \\ 
					2.5455   6.5822e-02 \\ 
					2.6364   6.3552e-02 \\ 
					2.7273   6.1434e-02 \\ 
					2.8182   5.9452e-02 \\ 
					2.9091   5.7594e-02 \\ 
					3.0000   5.5849e-02 \\ 
					3.0909   5.4206e-02 \\ 
					3.1818   5.2657e-02 \\ 
					3.2727   5.1195e-02 \\ 
					3.3636   7.6545e-02 \\ 
					3.4545   7.4531e-02 \\ 
					3.5455   7.2620e-02 \\ 
					3.6364   7.0804e-02 \\ 
					3.7273   6.9077e-02 \\ 
					3.8182   6.7433e-02 \\ 
					3.9091   6.5864e-02 \\ 
					4.0000   6.4368e-02 \\ 
					4.0909   6.2937e-02 \\ 
					4.1818   6.1569e-02 \\ 
					4.2727   6.0259e-02 \\ 
					4.3636   5.9004e-02 \\ 
					4.4545   7.6319e-02 \\ 
					4.5455   7.4792e-02 \\ 
					4.6364   7.3326e-02 \\ 
					4.7273   7.1916e-02 \\ 
					4.8182   7.0559e-02 \\ 
					4.9091   6.9252e-02 \\ 
					5.0000   6.7993e-02 \\ 
					5.0909   6.6779e-02 \\ 
					5.1818   6.5607e-02 \\ 
					5.2727   6.4476e-02 \\ 
					5.3636   6.3383e-02 \\ 
					5.4545   6.2327e-02 \\ 
					5.5455   6.1305e-02 \\ 
					5.6364   7.4209e-02 \\ 
					5.7273   7.3031e-02 \\ 
					5.8182   7.1890e-02 \\ 
					5.9091   7.0784e-02 \\ 
					6.0000   6.9711e-02 \\ 
					6.0909   6.8671e-02 \\ 
					6.1818   6.7661e-02 \\ 
					6.2727   6.6680e-02 \\ 
					6.3636   6.5728e-02 \\ 
					6.4545   6.4802e-02 \\ 
					6.5455   6.3902e-02 \\ 
					6.6364   6.3027e-02 \\ 
					6.7273   7.3398e-02 \\ 
					6.8182   7.2419e-02 \\ 
					6.9091   7.1466e-02 \\ 
					7.0000   7.0538e-02 \\ 
					7.0909   6.9634e-02 \\ 
					7.1818   6.8752e-02 \\ 
					7.2727   6.7893e-02 \\ 
					7.3636   6.7055e-02 \\ 
					7.4545   6.6237e-02 \\ 
					7.5455   6.5439e-02 \\ 
					7.6364   6.4660e-02 \\ 
					7.7273   6.3899e-02 \\ 
					7.8182   7.2550e-02 \\ 
					7.9091   7.1716e-02 \\ 
					8.0000   7.0901e-02 \\ 
					8.0909   7.0104e-02 \\ 
					8.1818   6.9325e-02 \\ 
					8.2727   6.8563e-02 \\ 
					8.3636   6.7818e-02 \\ 
					8.4545   6.7089e-02 \\ 
					8.5455   6.6375e-02 \\ 
					8.6364   6.5677e-02 \\ 
					8.7273   6.4992e-02 \\ 
					8.8182   6.4322e-02 \\ 
					8.9091   7.1731e-02 \\ 
					9.0000   7.1006e-02 \\ 
					9.0909   7.0296e-02 \\ 
					9.1818   6.9600e-02 \\ 
					9.2727   6.8918e-02 \\ 
					9.3636   6.8249e-02 \\ 
					9.4545   6.7592e-02 \\ 
					9.5455   6.6949e-02 \\ 
					9.6364   6.6317e-02 \\ 
					9.7273   6.5697e-02 \\ 
					9.8182   6.5089e-02 \\ 
					9.9091   6.4492e-02 \\ 
					10.0000   7.0962e-02 \\ 
				};
				\addlegendentry{PVTC scheme~\cite{Paolini2022IRSA_BAC}};
				
				\addplot [line width = 1.5,color=red]
				table[row sep=crcr]{%
					1.0000   2.5747e-01 \\ 
					1.0909   3.1163e-01 \\ 
					1.1818   2.8766e-01 \\ 
					1.2727   2.6711e-01 \\ 
					1.3636   2.4931e-01 \\ 
					1.4545   2.3373e-01 \\ 
					1.5455   2.1998e-01 \\ 
					1.6364   2.0776e-01 \\ 
					1.7273   1.9682e-01 \\ 
					1.8182   1.8698e-01 \\ 
					1.9091   1.7808e-01 \\ 
					2.0000   1.6998e-01 \\ 
					2.0909   1.6259e-01 \\ 
					2.1818   1.5582e-01 \\ 
					2.2727   1.8404e-01 \\ 
					2.3636   1.7696e-01 \\ 
					2.4545   1.7041e-01 \\ 
					2.5455   1.6432e-01 \\ 
					2.6364   1.5865e-01 \\ 
					2.7273   1.5336e-01 \\ 
					2.8182   1.4842e-01 \\ 
					2.9091   1.4378e-01 \\ 
					3.0000   1.3942e-01 \\ 
					3.0909   1.3532e-01 \\ 
					3.1818   1.3146e-01 \\ 
					3.2727   1.2780e-01 \\ 
					3.3636   1.2435e-01 \\ 
					3.4545   1.2108e-01 \\ 
					3.5455   1.1797e-01 \\ 
					3.6364   1.1502e-01 \\ 
					3.7273   1.1222e-01 \\ 
					3.8182   1.0955e-01 \\ 
					3.9091   1.0700e-01 \\ 
					4.0000   1.0457e-01 \\ 
					4.0909   1.0224e-01 \\ 
					4.1818   1.0002e-01 \\ 
					4.2727   9.7892e-02 \\ 
					4.3636   9.5853e-02 \\ 
					4.4545   1.1085e-01 \\ 
					4.5455   1.0863e-01 \\ 
					4.6364   1.0650e-01 \\ 
					4.7273   1.0445e-01 \\ 
					4.8182   1.0248e-01 \\ 
					4.9091   1.0058e-01 \\ 
					5.0000   9.8753e-02 \\ 
					5.0909   9.6990e-02 \\ 
					5.1818   9.5288e-02 \\ 
					5.2727   9.3645e-02 \\ 
					5.3636   9.2058e-02 \\ 
					5.4545   9.0524e-02 \\ 
					5.5455   8.9040e-02 \\ 
					5.6364   8.7603e-02 \\ 
					5.7273   8.6213e-02 \\ 
					5.8182   8.4866e-02 \\ 
					5.9091   8.3560e-02 \\ 
					6.0000   8.2294e-02 \\ 
					6.0909   8.1066e-02 \\ 
					6.1818   7.9874e-02 \\ 
					6.2727   7.8716e-02 \\ 
					6.3636   7.7592e-02 \\ 
					6.4545   7.6499e-02 \\ 
					6.5455   7.5436e-02 \\ 
					6.6364   7.4403e-02 \\ 
					6.7273   7.3398e-02 \\ 
					6.8182   7.2419e-02 \\ 
					6.9091   7.1466e-02 \\ 
					7.0000   7.0538e-02 \\ 
					7.0909   6.9634e-02 \\ 
					7.1818   6.8752e-02 \\ 
					7.2727   6.7893e-02 \\ 
					7.3636   6.7055e-02 \\ 
					7.4545   6.6237e-02 \\ 
					7.5455   6.5439e-02 \\ 
					7.6364   6.4660e-02 \\ 
					7.7273   6.3899e-02 \\ 
					7.8182   6.3156e-02 \\ 
					7.9091   6.2430e-02 \\ 
					8.0000   6.1721e-02 \\ 
					8.0909   6.1027e-02 \\ 
					8.1818   6.0349e-02 \\ 
					8.2727   5.9686e-02 \\ 
					8.3636   5.9037e-02 \\ 
					8.4545   5.8402e-02 \\ 
					8.5455   5.7781e-02 \\ 
					8.6364   5.7173e-02 \\ 
					8.7273   5.6577e-02 \\ 
					8.8182   5.5994e-02 \\ 
					8.9091   6.3666e-02 \\ 
					9.0000   6.3023e-02 \\ 
					9.0909   6.2393e-02 \\ 
					9.1818   6.1775e-02 \\ 
					9.2727   6.1169e-02 \\ 
					9.3636   6.0575e-02 \\ 
					9.4545   5.9993e-02 \\ 
					9.5455   5.9422e-02 \\ 
					9.6364   5.8861e-02 \\ 
					9.7273   5.8311e-02 \\ 
					9.8182   5.7771e-02 \\ 
					9.9091   5.7241e-02 \\ 
					10.0000   5.6721e-02 \\
				};
				\addlegendentry{Our scheme, ED-MPR \revise{(lower bound)}};
			\end{axis}
		\end{tikzpicture}
		\caption{The achievable average sum rate $\bar{R}_{\rm sum} = \frac{1-\delta}{\beta}G^*(\Lambda)$ with vanishing \gls{plr} in the $(G,D,\beta,\delta)$-asymptotic regime with $D= 1$, $\delta = 0.9$, $\nu = 0.5$, and $\Lambda(x) = x^2$.}
		\label{fig:sum_rate}
	\end{figure}

	Finally, we compare in Fig.~\ref{fig:sum_rate} the average sum rate $\bar{R}_{\rm sum} = \frac{1-\delta}{\beta}G^*(\Lambda)$ achievable with vanishing \gls{plr} in the $(G,D,\beta,\delta)$-asymptotic regime by the original IRSA, the PVTC scheme, and our scheme with ED-MPR. \revise{For our scheme, $G^*(\Lambda)$ is replaced by the lower bound in Theorem~\ref{th:MPR_asymptotic}.} We consider $\delta = 0.9$ and $\Lambda(x) = x^2$. As expected, the original IRSA with single-packet reception achieves the lowest sum rate. Our scheme significantly outperforms the PVTC scheme when $\beta$ is small. 
	
	\section{Conclusion} \label{sec:conclusion}
	We discussed an IRSA-based random-access scheme for the binary adder channel. In a slot, the active users transmit a pilot symbol followed by \gls{iid} binary symbols. We proposed a per-slot decoder that discards incompatible codewords from a list of candidate codewords, and an improved decoder that shrinks this list across iterations. Our scheme resolves more colliding users per slot and thus achieves a higher sum rate than the state-of-the-art scheme proposed in~\cite{Paolini2022IRSA_BAC}, in the regime where the per-user payload increases slowly with the number of users, and the total number of bits transmitted per channel use in a slot is close to~$1$. A subject for future work is the asymptotic analysis of the improved decoder.
	
	\appendices
%
	\section{Proof of Theorem~\ref{th:MPR_expurgation}} \label{app:MPR_expurgation}
	    Consider a degree-$U$ slot. Let $\Ac_0$ and $\Ac_U$ be the set of channel uses in this slot where 
		the channel output is~$0$ and $U$, respectively. Let $\rv{A}_0 = |\Ac_0|$ and $\rv{A}_U = |\Ac_U|$. After some manipulations, we can derive the joint \gls{pmf} of $\rv{A}_0$ and $\rv{A}_U$ as in~\eqref{eq:pmf_A0AU}.\footnote{We treat the codewords as independent although they are required to be distinct. The probability of generating two identical codewords is $(\nu^2 + (1-\nu)^2)^{n_0}$, which is negligible for large $n_0$.} 
		Let $\rv{L}_0$ be the number of nontransmitted codewords associated with the slot. The slot is resolved if none of these $\rv{L}_0$ nontransmitted codewords survive the discarding process. A nontransmitted codeword survives if its entries in all positions in $\Ac_0 \cup \Ac_U$ agree with the channel outputs, i.e., the entries in positions in $\Ac_U$ are $1$ and the entries in positions in $\Ac_0$ are~$0$. The probability of this event is $\nu^{\rv{A}_U} (1-\nu)^{\rv{A}_0}$. It follows that the probability $\pi_U$ that the slot is resolved is
		\begin{align}
			\pi_U =  \revise{\EE_{\rv{L}_0, \rv{A}_0, \rv{A}_U}\Big[\big(1 - \nu^{\rv{A}_U} (1-\nu)^{\rv{A}_0}\big)^{\rv{L}_0}\Big].} \label{eq:tmp215}
		\end{align}
		To compute the \gls{pmf} of $\rv{L}_0$, notice that a generic codeword is associated with $\Lambda'(1)$ slots in average. Thus the probability that a generic codeword is associated with a given slot is $\Lambda'(1)/N$. Therefore, the probability that $L$ out of $M-U$ nontransmitted codewords are associated with a given slot is given by
		\begin{align}
			\P[\rv{L}_0 = L] 
			&=  \binom{M-U}{L} \left(\frac{\Lambda'(1)}{N}\right)^L \left(1-\frac{\Lambda'(1)}{N}\right)^{M-U-L}. \notag
		\end{align}
		It follows that 
		\begin{align}
			\E[x^{\rv{L}_0}] &= \sum_L \binom{M-U}{L} \left(\frac{\Lambda'(1)}{N} x\right)^L \left(1-\frac{\Lambda'(1)}{N}\right)^{M-U-L} \notag \\
			&= \Big(1-\frac{\Lambda'(1)}{N} (1-x)\Big)^{M-U}. \label{eq:tmp225}
		\end{align}
		Next, by setting $x = 1 - \revise{\nu^{\rv{A}_U} (1-\nu)^{\rv{A}_0}}$ in~\eqref{eq:tmp225} and by using~\eqref{eq:tmp215}, we obtain~\eqref{eq:piU}. \revise{If $\nu = 0.5$, \eqref{eq:piU_nu0.5} follows directly from~\eqref{eq:piU} by noting that $\rv{A}_0 + \rv{A}_U \sim \Bino(n_0,2^{1-U})$.} 
        \revise{Finally, by applying Jensen's inequality, we lower-bound $\pi_U$ in~\eqref{eq:piU_nu0.5} by $\pi'_U = \big(1-\frac{\Lambda'(1)}{N} \EE\big[2^{-\rv{A}}\big]\big)^{M-U} = \big(1-\frac{\Lambda'(1)}{N} (1-2^{-U})^{n_0}\big)^{M-U}$.}
	
	\section{Proof of Theorem~\ref{th:MPR_asymptotic}} \label{app:MPR_asymptotic}
        \revise{We shall show that the lower bound $\pi'_U$ converges to $\bar{\pi}_U$ in the $(G,D,\beta,\delta)$-asymptotic regime.}
	    First, by applying the inequalities $\frac{z}{1+z} \le \ln(1+z) \le z, \forall z > - 1$, with $z=-\frac{\Lambda'(1)}{N}x$, and after some manipulations, we obtain that
		\begin{multline}
			\exp\!\bigg(\frac{-\frac{M-U}{N}\Lambda'(1)x}{1-\Lambda'(1)x/N}\bigg) \\ \le \Big(1-\frac{\Lambda'(1)}{N} x \Big)^{M-U} \le \exp\!\Big(-\frac{M-U}{N}\Lambda'(1)x\Big). \label{eq:tmp165}
		\end{multline}
		Next, let $\mu K/N = G$, $M = D K^{1/\delta}$, and $n_0 = \beta \log_2M$. 
		Setting $x = (1-2^{-U})^{n_0}$ in~\eqref{eq:tmp165}, we obtain
		\begin{align*}
			&\exp\!\bigg(\frac{-\frac{DK^{1/\delta}-U}{\mu K/G} \Lambda'(1) (1 - 2^{-U})^{\frac{\beta}{\delta}\log_2 K \revise{+ \beta \log_2D}}}{1- \Lambda'(1) (1 - 2^{-U})^{\frac{\beta}{\delta}\log_2 K \revise{+ \beta \log_2D}} G/(\mu K)}\bigg) \notag \\ 
			&~ \le \pi'_U \\
			&~ \le \exp\!\bigg(-\frac{DK^{1/\delta} -U}{\mu K/G} \Lambda'(1) (1 - 2^{-U})^{\frac{\beta}{\delta}\log_2 K \revise{+ \beta \log_2D}}\bigg).
		\end{align*}
		As $K \to\infty$, one can verify that both the lower bound and upper bound above converge to $\bar{\pi}_U$ given in~\eqref{eq:piU_asymptotic}.
        We conclude that  $\pi'_U \to \bar{\pi}_U$, \revise{and thus $\pi_U \ge \bar{\pi}_U$,} in the considered asymptotic regime. 
  
        It follows from Lemma~\ref{lemma:thres_MPR} in Appendix~\ref{app:IRSA_MPR} that \revise{for a scheme that resolves a collision of $U$ users with probability $\pi'_U$,} the decoding threshold is given by the largest value of $G$ such that~\eqref{eq:MPR_dec_thres} holds. \revise{Our scheme resolves collisions with higher probability, thus achieves a lower PLR, and in turn a higher decoding threshold.}
		
	\section{IRSA With Probabilistic \gls{MPR}} \label{app:IRSA_MPR}
	The existing analyses of IRSA with \gls{MPR}, such as~\cite{Ghanbarinejad2013Irregular}, assume that the receiver can resolve a collision with probability~$1$ if the number of colliding users is at most $T$, and the collision is unresolved otherwise. With degree distribution $\Lambda(x) = x$, this is also called $T$-fold slotted ALOHA~\cite{PolyanskiyISIT2017massive_random_access}. In this appendix, we extend the decoding threshold analysis of IRSA with \gls{MPR} in~\cite{Ghanbarinejad2013Irregular} to a more general setting where a collision of~$U$ users is resolved with probability $\pi_U\of{N}$ that depends on the number of slots per frame, $N$. \revise{This result can be of independent interest.}
	
	
	\begin{lemma} \label{lemma:thres_MPR}
		Consider IRSA with $N$ slots per frame, degree distribution $\Lambda(x)$, and \gls{MPR} capability where a collision of $U$ users is resolved with probability $\pi_U\of{N}$. Assume that $\pi_U\of{N} \to \tilde{\pi}_U$ as $N \to\infty$. The decoding threshold $G^*$ of this scheme is the largest value of $G$ such that
		\begin{equation} \label{eq:thres_MPR}
			p > \lambda\bigg( 1 - e^{-G \Lambda'(1)p} \sum_{u=0}^{T-1} \frac{\tilde{\pi}_{u+1}}{u!} (p G \Lambda'(1))^u \bigg),~~ \forall p \in (0,1]. 
		\end{equation}
		where $\lambda(x) = \Lambda'(x)/\Lambda'(1)$ and $T = \max \{u \colon \tilde{\pi}_u > 0\}$. 
	\end{lemma}
	\begin{IEEEproof}
		Consider a degree-$L$ VN. Let $p$ be the probability that an edge is unknown, given that each of the other $L-1$ edges has been revealed with probability $1-q$ in the previous step. Since the edge is revealed whenever at least one of the other edges have been revealed, we have that $p = q^{L-1}$. Next, consider a degree-$L$ CN. Here, $q$ denotes the probability that an edge is unknown, given that each of the other $L-1$ edges has been revealed with probability $1-p$ in the previous step. By assumption, the edge is revealed with probability $\pi_{u+1}\of{N}$ if the number of unrevealed edges among the other $L-1$ edges is $u$. Therefore, 
			$1-q = \sum_{u = 0}^{L-1} \pi_{u+1}\of{N} \binom{L-1}{u} p^u (1-p)^{L-u-1}$.
		According to a tree analysis argument, by averaging the expressions of $p$ and $q$ over the edge distributions, we can derive the evolution of the average erasure probabilities during the $i$-th iteration as 
		\begin{align}
			p_i &= \sum_{L} \lambda_L q_{i-1}^{L-1} = \lambda(q_{i-1}), \label{eq:tmp323} \\
			q_i &= \sum_{L} \rho_L \bigg(1-\sum_{u = 0}^{L-1} \pi\of{N}_{u+1} \binom{L-1}{u} p_i^u (1-p_i)^{L-u-1} \bigg) \notag \\
			&= 1 - \sum_{u \colon \pi\of{N}_{u+1} > 0} \frac{\pi\of{N}_{u+1}}{u!} p_{i}^u \rho\of{u}(1-p_i),
		\end{align}
		where $\rho_L$ is the probability that an edge is connected to a degree-$L$ CN, $\rho(x) = \sum_L \rho_L x^{L-1}$, and $\rho\of{u}(x)$ is the $u$th order derivative of $\rho(x)$. As $N\to\infty$, $\rho(x) = e^{-G \Lambda'(1)(1-x)}$, and it follows that $\rho\of{u}(x) = (G \Lambda'(1))^u e^{-G \Lambda'(1)(1-x)}$. Furthermore, $\pi_{u+1}\of{N} \to \tilde{\pi}_{u+1}$. Therefore, with $T = \max\{u: \tilde{\pi}_u > 0\}$,
		\begin{align}
			q_i &= 1 - \sum_{u=0}^{T-1} \frac{\tilde{\pi}_{u+1}}{u!} p_{i}^u (G \Lambda'(1))^u e^{-G \Lambda'(1)p_i} \\
			&= 1 - e^{-G \Lambda'(1)p_i} \sum_{u=0}^{T-1} \frac{\tilde{\pi}_{u+1}}{u!} (p_{i} G \Lambda'(1))^u. \label{eq:tmp333}
		\end{align}
		By combining~\eqref{eq:tmp323} and~\eqref{eq:tmp333}, we obtain
		\begin{align}
			p_i = \lambda\bigg(1 - e^{-G \Lambda'(1)p_{i-1}} \sum_{u=0}^{T-1} \frac{\tilde{\pi}_{u+1}}{u!} (p_{i-1} G \Lambda'(1))^u\bigg).
		\end{align}
		The decoding threshold $G^*$ is the largest value of $G$ such that $\{p_i\}$ is a decreasing sequence, i.e., $p_{i-1} > p_i$ for all $i$. The inequality~\eqref{eq:thres_MPR} ensures this property.
	\end{IEEEproof}
	
	\bibliographystyle{IEEEtran}
	\bibliography{IEEEabrv,./biblio}
\end{document}